\documentclass[fleqn,usenatbib]{mnras}

\usepackage{newtxtext,newtxmath}

\usepackage[T1]{fontenc}

\DeclareRobustCommand{\VAN}[3]{#2}
\let\VANthebibliography\thebibliography
\def\thebibliography{\DeclareRobustCommand{\VAN}[3]{##3}\VANthebibliography}

\usepackage{graphicx}	
\usepackage{amsmath}	

\newcommand{\msol}{$M_{\odot}$}

\title[Simulated optical light curves of ZEBRAs]{Simulated optical light curves of super-Eddington tidal disruption events with ZEBRA flows}

\author[R. A. J. Eyles-Ferris et al.]{
R. A. J. Eyles-Ferris,$^{1}$\thanks{raje1@leicester.ac.uk}
R. L. C. Starling,$^{1}$
P. T. O'Brien,$^{1}$
C. J. Nixon,$^{1}$
and Eric R.~Coughlin$^{2}$
\\
$^{1}$School of Physics and Astronomy, University of Leicester, University Road, Leicester, LE1 7RH, UK\\
$^{2}$Department of Physics, Syracuse University, Syracuse, NY 13210, USA\\
}

\date{Accepted XXX. Received YYY; in original form ZZZ}

\pubyear{2022}

\begin{document}
\label{firstpage}
\pagerange{\pageref{firstpage}--\pageref{lastpage}}
\maketitle

\defcitealias{Coughlin&Begelman}{CB14}
\defcitealias{Wu18}{WCN18}

\begin{abstract}
We present simulated optical light curves of super-Eddington tidal disruption events (TDEs) using the zero-Bernoulli accretion (ZEBRA) flow model, which proposes that during the super-Eddington phase, the disc is quasi-spherical, radiation-pressure dominated, and accompanied by the production of strong jets. We construct light curves for both on- and off-axis (with respect to the jet) observers to account for the anisotropic nature of the jetted emission. We find that at optical wavelengths, emission from the accretion flow is orders of magnitude brighter than that produced by the jet, even with boosting from synchrotron self-Compton. Comparing to the observed jetted TDE Swift J2058.4+0516, we find that the ZEBRA model accurately captures the timescale for which accretion remains super-Eddington and reproduces the luminosity of the transient. However, we find the shape of the light curves deviate at early times and the radius and temperature of our modelled ZEBRA are $\sim2.7 - 4.1$ times smaller and $\sim1.4 - 2.3$ times larger, respectively, than observed. We suggest that this indicates the ZEBRA inflates more, and more rapidly, than currently predicted by the model, and we discuss possible extensions to the model to account for this. Such refinements, coupled with valuable new data from upcoming large scale surveys, could help to resolve the nature of super-Eddington TDEs and how they are powered.
\end{abstract}




\section{Introduction}

When a star with radius $R_*$ and mass $M_*$ comes to within the tidal radius 
\begin{equation}
    r_t \simeq R_*(M_{\text{BH}}/M_*)^{1/3}
\end{equation}
of a super massive black hole (SMBH) with mass $M_{\text{BH}}$, the extreme tidal force of the SMBH overcomes the self-gravity of the star, producing a tidal disruption event \citep[TDE; e.g.,][]{Rees88,Gezari21}. Approximately half the disrupted material remains bound to the SMBH and falls back onto it to be accreted, with the rate varying approximately as $\propto t^{-5/3}$ when the star is completely destroyed \citep{Phinney89}, or as $\propto t^{-9/4}$ if the disruption is partial and the core of the star survives the encounter \citep{Coughlin19, Miles20, Nixon21}.

At early times, the rate at which material falls back to the SMBH, the fallback rate $\dot{M}_{\rm fb}$, can be much higher than the Eddington rate of the SMBH. As the accretion rate, $\dot{M}_{\rm acc}$, is assumed to be comparable to $\dot{M}_{\rm fb}$, the SMBH undergoes super-Eddington accretion. Several interpretations of such super-Eddington TDEs have been proposed. Both \citet{Strubbe09} and \citet{Lodato11} presented models in which the extreme radiation pressure causes sub-relativistic outflows to be launched (see also \citealt{Metzger16}). Such outflows emit from their photospheres as blackbodies \citep{Loeb97}, however, these models predict significantly different temperatures from what has been observed \citep[e.g.][]{vanVelzen11,Pasham15}.

More recently, models have been proposed that combine both exotic accretion flows and outflows, and here we utilise one such model, the `ZEro-BeRnoulli Accretion" (ZEBRA) flow model developed by \citet[hereafter \citetalias{Coughlin&Begelman}]{Coughlin&Begelman}. To briefly summarise, at early times the angular velocity of the material falling back to the SMBH is too low to prevent super-Eddington accretion rates. The low angular momentum, in addition to the inefficient advection of accretion energy, results in a puffed-up, quasi-spherical accretion flow -- the ZEBRA -- instead of a thin disk. Reprocessing of the super-Eddington accretion luminosity within the envelope leads to thermal emission from the surface of the ZEBRA flow at the Eddington limit, where photons are no longer efficiently trapped \citep{Begelman78}. Inflated `disks' are also proposed in other models, such as that of \citet{Dai18} and \citet{Metzger22}.  \citeauthor{Metzger22}'s model assumes that the debris stream rapidly circularises to form a quasi-spherical `disk' with Eddington-limited emission similar to the ZEBRA model. \citeauthor{Dai18}, on the other hand, started with initial conditions that resembled a thick disc (H/R $\sim$ 0.3), though it is not clear if the total mass and angular momentum (which they assumed to be Keplerian) were representative of those in a TDE, and thus how accurately their flow would map to a ZEBRA. The presence of inflated `disks' has also been identified numerically with several simulations found to be consistent with the predictions of the ZEBRA model \citep{Sadowski16,Bonnerot20,Andalman22}.

However, the ZEBRA flow itself can only radiate at the Eddington limit of the SMBH and remain quasi-hydrostatic. The remaining excess accretion energy must therefore be exhausted, and the model proposes that this occurs through the launching of relativistic jets that propagate along the (nearly evacuated) poles of the flow. The model of \citet{Dai18} also includes such jets, and it is posited the spin of the SMBH in the disk's magnetic field induces the Blandford-Znajek process and launches similar jets \citep{Blandford77}. The anisotropic nature of \citeauthor{Dai18}'s jets is suggested to contribute to the differences between X-ray and optical TDEs, with a strong angular dependence on the observed properties of a TDE. The \citet{Metzger22} model differs, however, in that super-Eddington fallback rates do not result in super-Eddington heating of the flow \citep[e.g.][]{Steinberg22} and therefore the removal of excess energy by jets or outflows is not necessary. \citeauthor{Dai18}, in addition to other authors \citep[e.g.][]{Metzger16,Gezari21}, suggest that super-Eddington accretion also results in sub-relativistic and wide-angle outflows. Such wide-angle outflows are, however, currently excluded in the ZEBRA model.

Jetted super-Eddington TDEs are believed to have already been observed. In particular, the \textit{Neil Gehrels Swift Observatory}, hereafter \textit{Swift}, has observed three likely super-Eddington TDEs, Swift J164449.3+573451 \citep[hereafter J1644+57,][]{Burrows11,Levan11,Zauderer11}, Swift J2058.4+0516 \citep[hereafter J2058+05,][]{Cenko12,Pasham15} and Swift J1112.2-8238 \citep[hereafter J1112-82,][]{Brown15}. All of these events exhibited properties consistent with a relativistic jet rather than a purely thermal transient. For instance, power law rather than thermal X-ray spectra were observed in all three cases, while significant IR and optical polarisation was identified for J1644+57 and J2058+05, respectively, which are consistent with jet contributions at these wavelengths \citep{Wiersema12,Wiersema20}. The erratic nature of the early X-ray light curve of J1644+57 was also shown to be consistent with jet wobbling due to a `magnetically arrested disk' \citep{Tchekohovskoy14}, but whether sufficient magnetic flux is present to form such an accretion flow remains to be demonstrated.

To investigate the geometric and accretion properties of ZEBRA flows, \citet[hereafter \citetalias{Wu18}]{Wu18} performed a series of simulations. Starting from the initial disruption of the star, they used the numerically calculated fallback rates to derive the shape and behaviour of a forming ZEBRA envelope across a period of three years. We use the results of these simulations to inform the properties of the ZEBRA envelope and derive light curves.

In Sections \ref{sec:ZEBRA_description} and \ref{sec:Wu_simulations}, we summarise the physics of ZEBRA flows and the numerical simulations of \citetalias{Wu18}. The results of those simulations are used to generate our light curves as detailed in Section \ref{sec:LC_description}. Sections \ref{sec:Results} and \ref{sec:Discussion} detail our results and compare them to observations. We present our conclusions in Section \ref{sec:Conclusions}.

Throughout this paper we adopt a flat $\Lambda$CDM cosmology with $H_0 =  71$ km\,s$^{-1}$\,Mpc$^{-1}$, $\Omega_m = 0.27$ and $\Omega_\Lambda = 0.73$.

\section{The zero-Bernoulli accretion (ZEBRA) model of tidal disruption events}
\label{sec:ZEBRA_description}

\subsection{Super-Eddington phase}

\begin{figure}
 \includegraphics[width=\columnwidth]{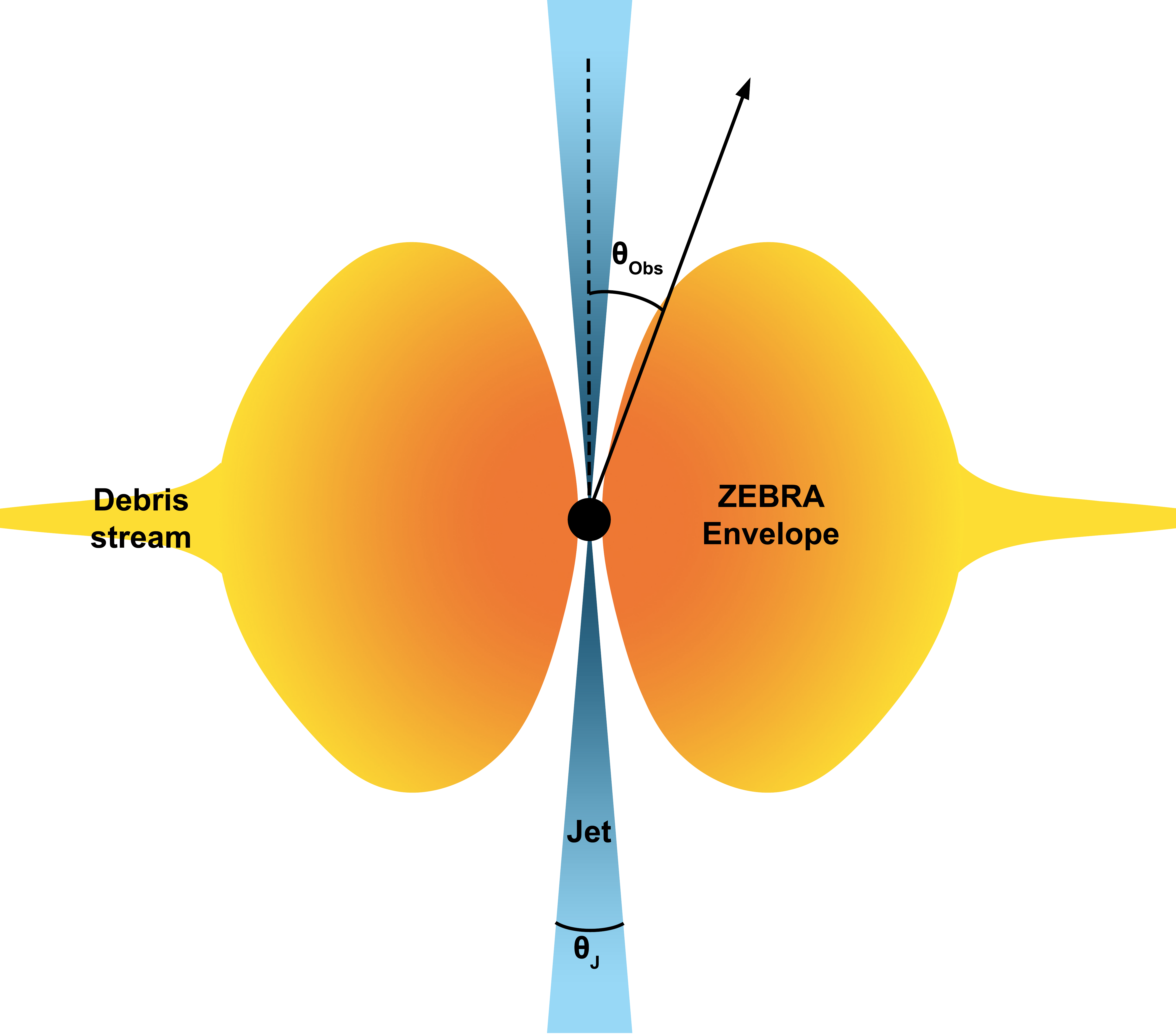}
 \caption{A sketch of the ZEBRA flow model. The opening angle of the jets produced due to super-Eddington accretion, $\theta_J$, is small. The visibility of the jet therefore depends on the angle of the observer from the poles of the ZEBRA envelope, $\theta_{\rm Obs}$.}
 \label{fig:ZEBRA_diagram}
\end{figure}

During the super-Eddington phase of the TDE, a zero-Bernoulli accretion (ZEBRA) flow is assumed to form. This is a special case of an ``adiabatic inflow-outflow solution'' (ADIOS) for which the Bernoulli parameter, $B$, is globally zero. The flow is not radiative and instead energy is assumed to be advectively transported throughout the flow and the material cannot cool efficiently. The combination of this inefficient cooling, the super-Eddington accretion rate and the low specific angular momentum of the gas implies that the flow is geometrically thick. The accretion and shocks within the flow also inject additional thermal energy, and coupled to the already-low binding energy of the returning material, result in the material being very weakly bound. However, rather than unbinding the disk, excess accretion energy is exhausted in the form of jets at the poles of the flow, as shown schematically in Figure \ref{fig:ZEBRA_diagram}.

The density, pressure and angular momentum squared are assumed to vary self-similarly with radius within the ZEBRA envelope, from some inner radius $r_0$, near the innermost stable circular orbit of the gas, to the trapping radius \citep{Begelman78}. These parameters vary as \citepalias{Coughlin&Begelman}:
\begin{equation}
\label{eq:ZEBRA_density}
\rho(r,\theta) = \rho_0\left(\frac{r}{r_0}\right)^{-q}(\sin^2{\theta})^{\alpha},
\end{equation}
\begin{equation}
\label{eq:ZEBRA_pressure}
p(r,\theta) = \beta \frac{G {M_{\rm BH}}}{r} \rho = \beta \frac{G {M_{\rm BH}}}{r} \rho_0\left(\frac{r}{r_0}\right)^{-q}(\sin^2{\theta})^{\alpha},
\end{equation}
and
\begin{equation}
\label{eq:ZEBRA_specificl}
\ell^2(r,\theta) = a G {M_{\rm BH}} r \sin^2{\theta},
\end{equation}
where constants $\alpha$, $\beta$ and $a$ obey
\begin{equation}
\label{eq:ZEBRA_alphabetaa}
\alpha = \frac{1 - q(\gamma - 1)}{\gamma - 1}, \quad
\beta = \frac{\gamma - 1}{1 + \gamma - q(\gamma - 1)}, \quad
a = 2 \alpha \beta
\end{equation}
in which $\gamma$ is the adiabatic index of the gas, taken to be $4/3$ as the accretion flow is dominated by radiation pressure (note also that $\beta$ in the equations above is not the impact parameter used to describe the ratio of tidal radius to the pericentre radius of the disrupted star).

The density power-law index $q$ can vary from 0.5 to 3, with the flow attaining a more spherical geometry as $q$ increases. The limits are set so that energy generation decreases with increasing radius ($q>0.5$) and to ensure the envelope does not become fully spherically symmetrical and a non-infinite density is maintained at the poles ($q<3$). For TDEs, $q$ can be determined from the bulk properties of the flow. In particular, rearranging the above equations yields the function $f$ which has two equivalent values and is dependent on the total mass, $\mathscr{M}=\int \rho dV$, angular momentum, $\mathscr{L}=\int \ell \rho dV$, and $M_{\rm BH}$:
\begin{equation}
\begin{split}
\label{eq:q}
f(\mathscr{M},\mathscr{L},M_{\rm BH}) 
&\equiv \left(\frac{y\kappa}{4\pi c}\right)^{1/6}\frac{\mathscr{M}\sqrt{G M_{\rm BH}}}{\mathscr{L}^{5/6}} \\[10pt]
&= \frac{\Gamma(\alpha+1)^{5/6}\Gamma(\alpha+2)^{5/6}}{\beta^{1/6}\alpha^{1/2}\Gamma(\alpha+3/2)^{5/3}}\frac{(7/2-q)^{5/6}}{3-q}
\end{split}
\end{equation}
where $\Gamma$ is the generalised factorial, $y\sim1$ is the efficiency of convection and $\kappa$ is the relevant opacity, in this case taken to be the opacity of electron scattering ($\kappa \sim 0.34$ cm$^2$ g$^{-1}$ for standard metallicities). As $y$ only enters Equation \eqref{eq:q} to the $1/6$ power, it has negligible impact on the solution for $q$. Using Equations \eqref{eq:ZEBRA_density} -- \eqref{eq:q}, the parameters of the envelope can therefore be determined at single instants in time and their evolution determined.

\citetalias{Coughlin&Begelman} introduced the factors $\delta$, $\chi$, and $y$ to parameterize the efficiency of accretion, the inner radius of the disc in terms of the Schwarzschild radius, and convective energy transport throughout the disc, respectively, and are not able to be further constrained within the model but are likely of the order ~0.1 - 1 for a geometrically thick ZEBRA flow (see the discussion above Equation 28 in \citetalias{Coughlin&Begelman}). The accretion rate can therefore be approximated as that of the spherically symmetric regime, $\dot{M}_{\rm acc} = 4 \pi r_0^2 \rho v_{r_0}$ where $v_{r_0}$ is the radial velocity at the inner radius of the ZEBRA envelope. This can be solved to give
\begin{equation}
\label{eq:ZEBRA_mdot_acc}
\begin{split}
\dot{M}_{\rm acc} = & \delta \chi^{3/2 - q} \mathscr{M}\sqrt{G M_{\rm BH}} \left(\frac{2 G M_{\rm BH}}{c^2}\right)^{3/2 - q} \\
& \times \left(\frac{y\sigma_{\rm T}}{4 \pi c m_p}\mathscr{M}\sqrt{G M_{\rm BH}}\right)^{-\frac{2}{5}(3-q)} h(q)   
\end{split}
\end{equation}
where $\delta$ is a number less than 1, $\chi = r_0 / R_S$ where $R_S$ is the Schwarzschild radius of the SMBH, $\sigma_{\rm T}$ is the Thomson scattering cross section and
\begin{equation}
\label{eq:ZEBRA_h}
h(q) \equiv \frac{2}{\sqrt{\pi}}\frac{\Gamma(\alpha+3/2)(3-q)}{\Gamma(\alpha+1)}((3-q)\beta\sqrt{a})^{-\frac{2}{5}(3-q)}.
\end{equation}
The photospheric radius can also be solved for, finding it to be coincident with the trapping radius:
\begin{equation}
\label{eq:ZEBRA_R}
\mathscr{R} = \left(\frac{y\kappa\beta\sqrt{a}(3-q)}{4\pi c}\mathscr{M}\sqrt{G M_{\rm BH}}\right)^{2/5}.
\end{equation}

From these equations, the emission properties of the envelope can be determined. As ZEBRAs radiate as blackbodies with a fixed bolometric luminosity equal to the Eddington luminosity of the central SMBH, the temperature of the photosphere can be inferred from the radius. Equations \eqref{eq:ZEBRA_density} -- \eqref{eq:ZEBRA_R} can be solved numerically for a given instant to derive the radius and temperature and characterise the emission of the ZEBRA.

However, as the bolometric luminosity is equal to the Eddington luminosity of the SMBH, the super-Eddington accretion means excess energy is continually injected into the inner regions of the ZEBRA. This energy cannot be efficiently advected to the surface of the envelope, does not form wide angle outflows as in the ADIOS case and its absorption into the envelope would result in a positive Bernoulli parameter and the entire flow becoming unbound. Instead, jets remove excess energy along the centrifugally supported funnels of the envelope, and have a total isotropic luminosity of
\begin{equation}
   L_{j, {\rm iso}} = \max\left\{\epsilon {\dot M}_{\rm acc}c^2 - L_{\rm Edd}, 0\right\},
\end{equation}
where $\epsilon = 0.1$ is the (assumed) accretion efficiency. Although it is expected that not all TDEs result in jets, as borne out by radio observations \citep[e.g.][]{vanVelzen13}, the anisotropic nature of the jet formation could be a contributory factor to the differences between jetted and apparently non-jetted TDEs. In some cases, it is plausible that jets do form but are only observable across a narrow range of observer angles ($\theta_{\rm Obs}$ in Figure \ref{fig:ZEBRA_diagram}) at optical and higher frequencies.

\subsection{Sub-Eddington phase}

As the fallback rate drops with time, $\dot{M}_{\rm acc}$ falls below the Eddington limit and the excess accretion energy ceases to be injected into the inner regions of the ZEBRA envelope. Without this radiation support, the ZEBRA will collapse. It is currently unclear exactly how this collapse occurs, but it is likely that the ZEBRA will collapse from its photosphere towards the SMBH to form a thin disk. Modelling this precisely is beyond the scope of this work and we have therefore used a fading ZEBRA component and rising disk component to approximate this behaviour in our light curves as discussed in Section \ref{sec:collapse}.

We use the approach of \citet{Lodato11} to model the thin disk, for which the properties of the disk are derived from $\dot{M}_{\rm acc}$, $M_{\rm BH}$ and the radius of closest approach, $r_p$. It should be noted that the dependence on $r_p$ assumed by \citeauthor{Lodato11} is incorrect for $\beta>1$ \citep[see e.g.][]{Norman21} but is applicable to our simulations where $\beta = 1$. The viscous time scale of the thin disk can have a significant impact on the accretion rate but at early times in the sub-Eddington phase, it is much smaller than the fallback timescale \citep{Cannizzo90,Lodato11}. The accretion rate is therefore dominated by the fallback rate. The viscous timescale does start to dominate at later times but on the order of decades \citep{Cannizzo90,Ulmer99} and we do not discuss this phase here.

\section{Numerical simulations}
\label{sec:Wu_simulations}

We derive our light curves from the results of the numerical simulations performed by \citetalias{Wu18} and we discuss these briefly here. \citetalias{Wu18} used the Smoothed Particle Hydrodynamics (SPH) code \textsc{phantom} \citep{Price17} to simulate the disruption of a star of $R_* = R_{\odot}$ and $M_* = M_{\odot}$ by SMBHs with masses $1\times 10^5$, $5\times 10^5$, $1\times 10^6$, $5\times 10^6$ and $1\times 10^7$ \msol. 

A polytrope with polytropic index $\gamma = 5/3$ was used to model the star with $10^7$ particles positioned to approximate the density distribution. The polytrope was also relaxed in isolation to smooth numerically-induced perturbations. It was then placed on a parabolic orbit with impact parameter $\beta \equiv r_t/r_p = 1$, where $r_p$ is the radius of closest approach.

Following the disruption of the star, the resultant debris stream was assumed to evolve adiabatically with an adiabatic index $\gamma = 5/3$, a reasonable assumption for the evolution of the debris at early times \citep{Coughlin16b}. At later times the effects of magnetic fields and radiative recombinations could have a significant impact on the debris stream's behaviour, however, these are assumed to be negligible here \citep{Kasen10,Bonnerot17,Guillochon17b}. The self-gravity of the debris stream was simulated using a bisective tree algorithm with an opening angle criterion \citep{Gafton11}.

To model the SMBH, a Newtonian point mass was used. An ``accretion radius'' was also initialised during the time where the debris stream extends to high radii and prior to the most bound part returning to $r_p$. The accretion radius was set to $3 r_t$ and particles that approached the SMBH to within this radius were ``accreted'' and removed from the simulation. The rate of this ``accretion'' was used to derive $\dot{M}_{\rm fb}$.

$\dot{M}_{\rm fb}$ was used to define the mass contained in the ZEBRA envelope through $\dot{\mathscr{M}} = \dot{M}_{\rm fb} - \dot{M}_{\rm acc}$. The total angular momentum of the envelope, $\mathscr{L}$, is similarly derived from the angular momentum of the debris, $\mathscr{L}_{\rm fb}$, which is dependent on the fallback rate and the star's initial angular momentum. As the debris falls onto the SMBH, conservation of angular momentum means it loses its angular momentum to the system and therefore $\mathscr{L}=\mathscr{L}_{\rm fb}$.

\begin{figure*}
    \centering
    \includegraphics[width=\columnwidth]{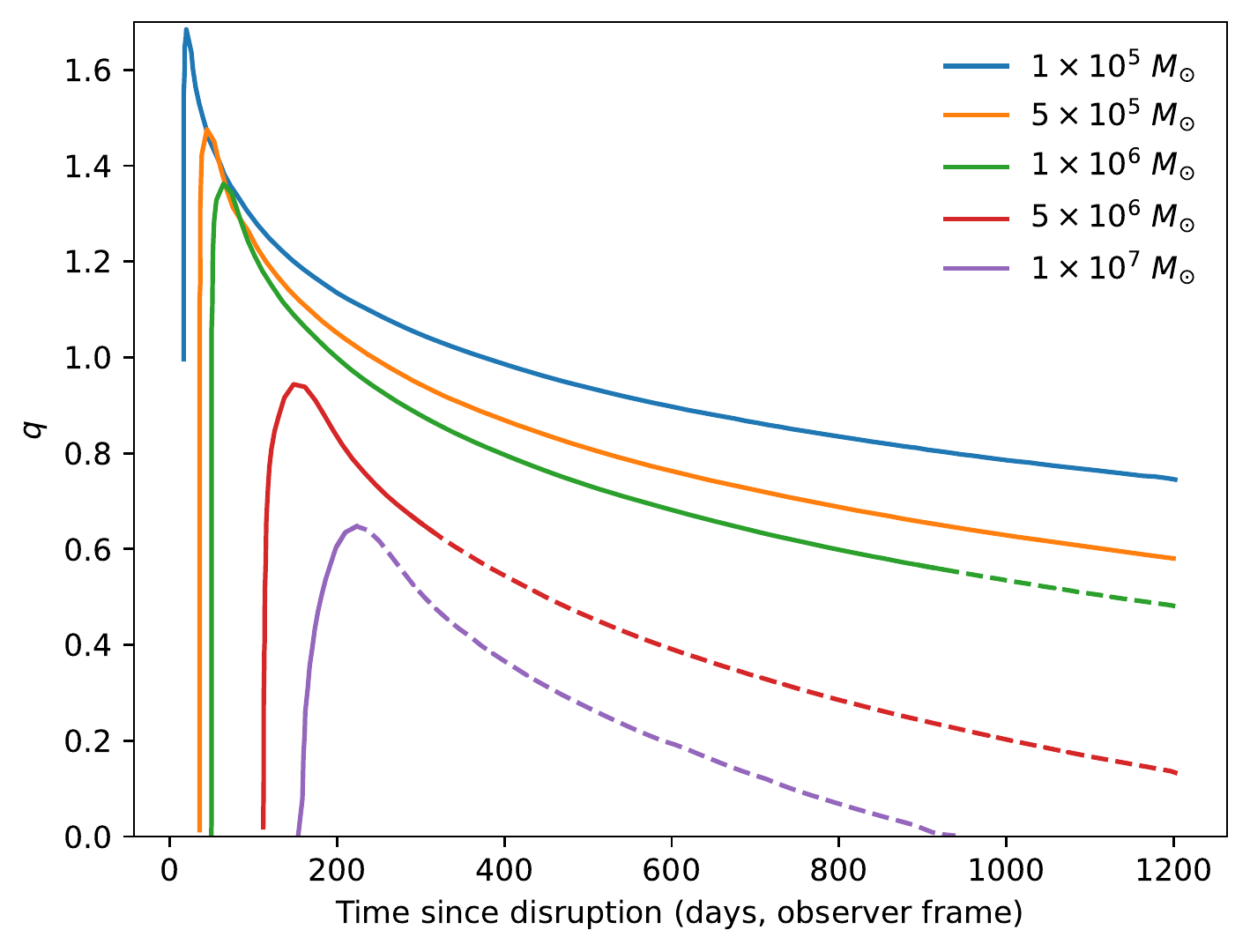}
    \includegraphics[width=\columnwidth]{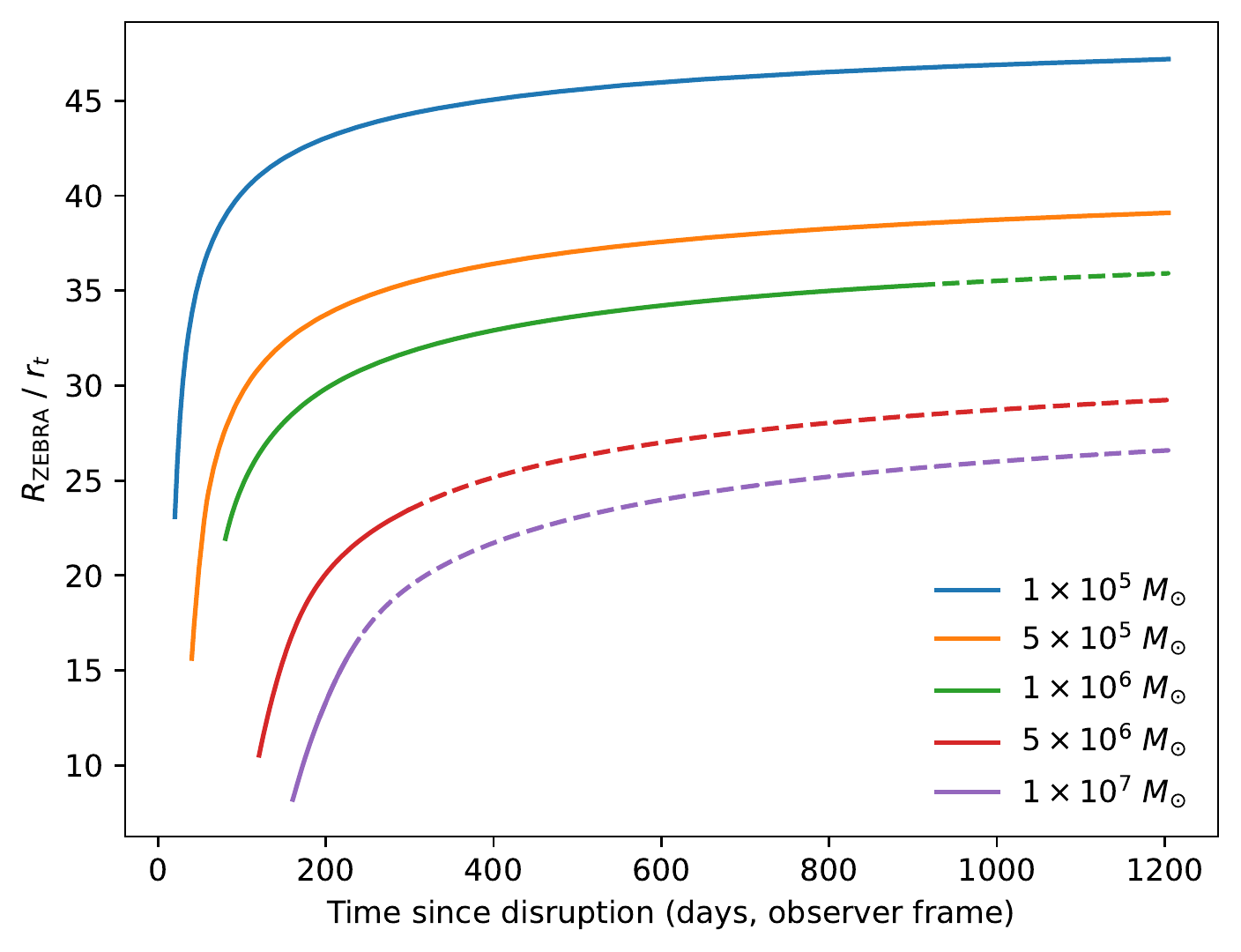}
    \includegraphics[width=\columnwidth]{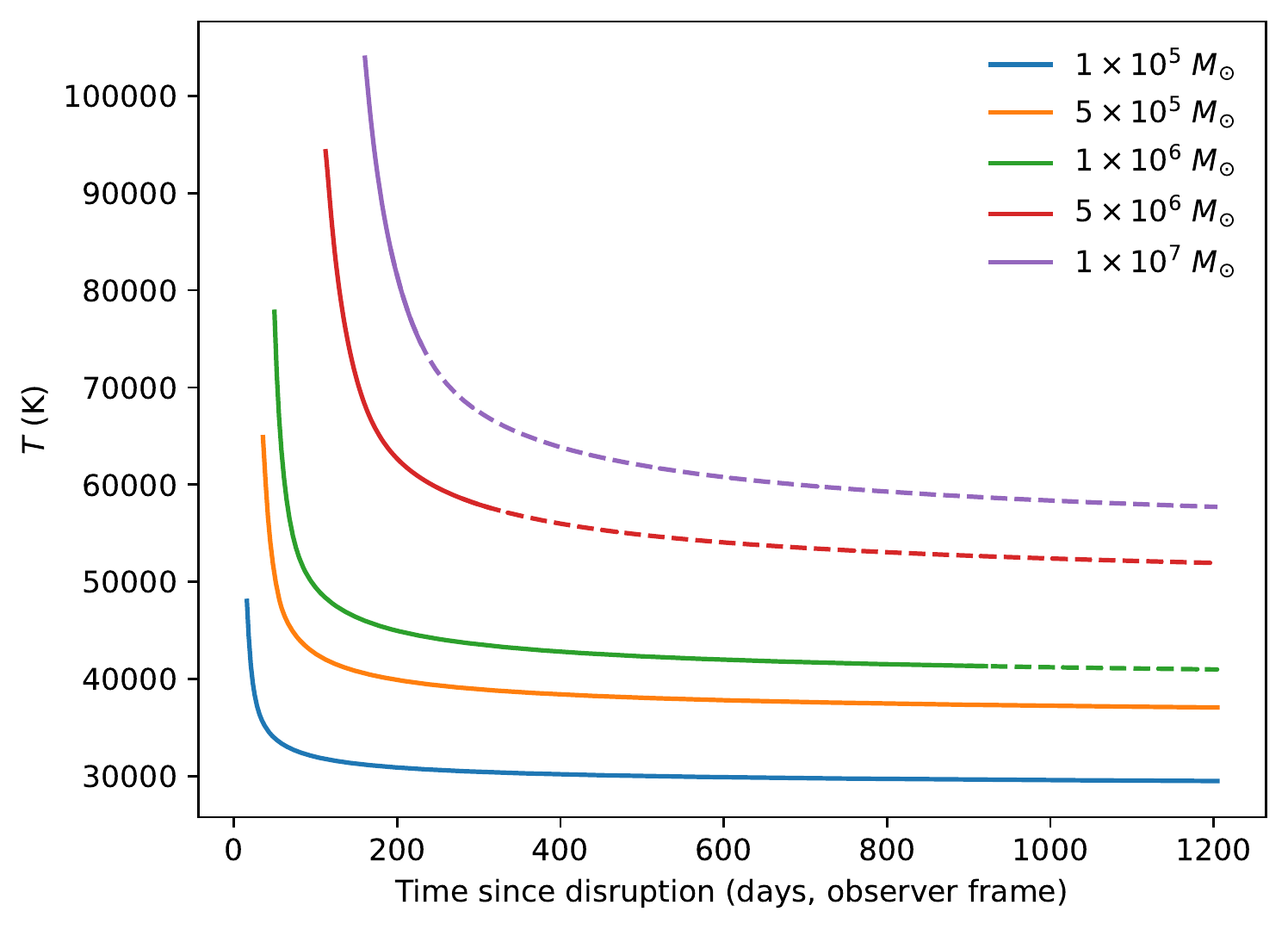}
    \includegraphics[width=\columnwidth]{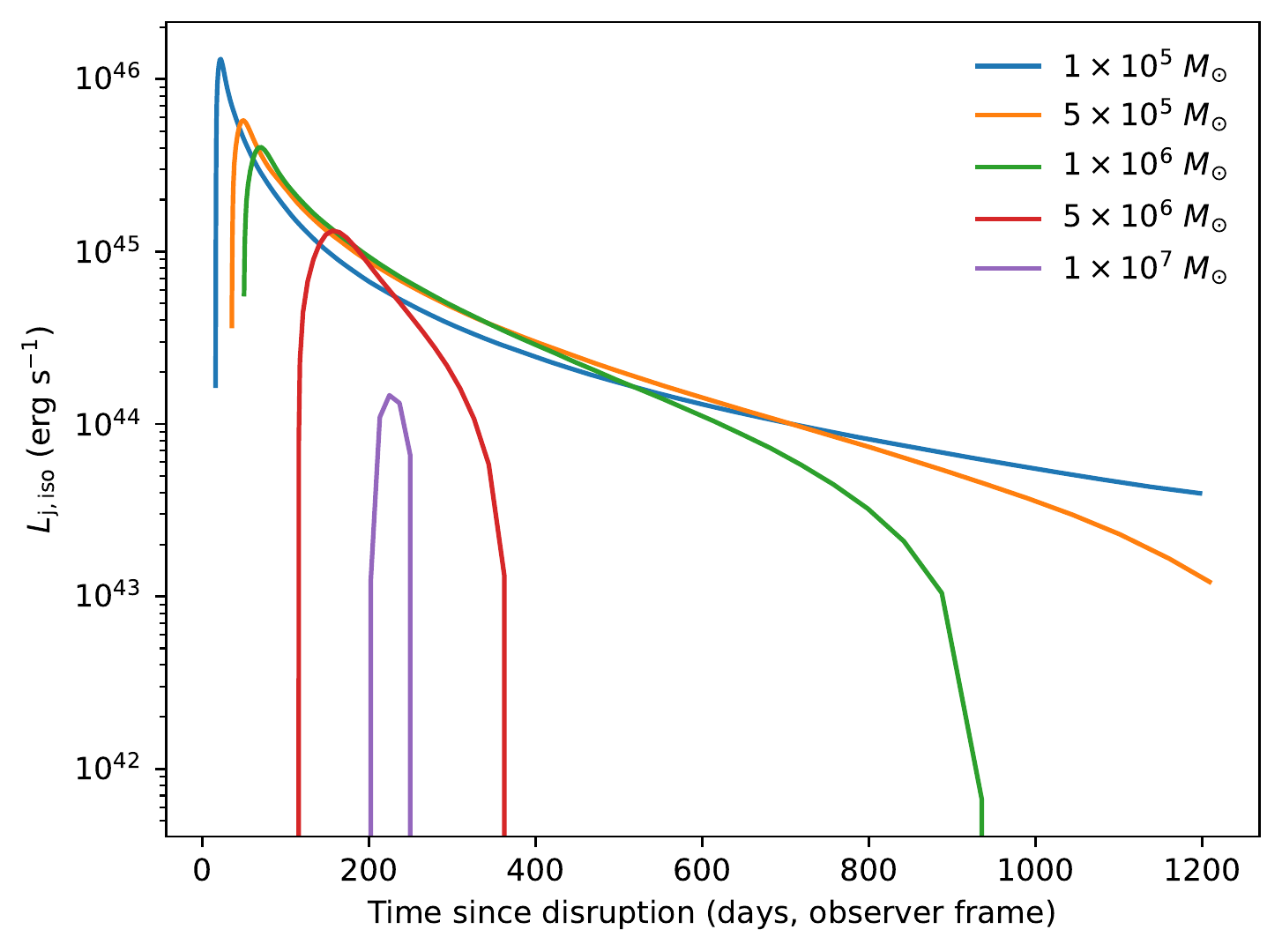}
    \caption{The results of \citetalias{Wu18}'s simulations extrapolated to TDEs occuring at $z=0.1$. Upper left: $q$; upper right: the radius of the ZEBRA's photosphere in units of $r_t$; lower left: the temperature of the ZEBRA's photosphere; and lower right: the total isotropic kinetic luminosity of the jets. In the all panels, the solid lines indicate times where the accretion rate was super-Eddington and in the first three panels, the dashed lines indicate a sub-Eddington accretion rate.}
    \label{fig:wcn_results}
\end{figure*}

From the derived $\mathscr{M}$ and $\mathscr{L}$, Equations \ref{eq:q} -- \ref{eq:ZEBRA_h} were used to numerically derive $q(t)$ from an initial $q$, although the solutions quickly converge to a solution that is independent of the initial value \citepalias{Coughlin&Begelman}. This also means the very earliest behaviour of the ZEBRA may not be accurately captured. From $q(t)$, $\dot{M}_{\rm acc}$ can be derived and the properties of the ZEBRA envelope can be determined, in particular, the photospheric radius and temperature and therefore the emission of the ZEBRA. The timescales for which $\dot{M}_{\rm acc}$ drops below the Eddington limit and the ZEBRA starts to collapse can also be calculated. The results of the simulations are summarised in Figure \ref{fig:wcn_results}. Note that these are for a source observed at $z=0.1$ for consistency with Figure \ref{fig:gband}.

\section{Light curve model}
\label{sec:LC_description}

In this Section, we discuss the emission of the ZEBRA and the derivation of our light curves. This emission consists of two main components, thermal radiation from the accretion flow, discussed in Section \ref{sec:accretionflow_emission}, and the contribution from the relativistic jet, examined in Section \ref{sec:jet_emission}. All light curves were calculated in the rest frame and two possibilities were examined where the observer was either on-axis with the jet ($\theta_{\rm Obs} = 0$\degr) or off-axis ($\theta_{\rm Obs} = 30$\degr). Any jet contribution was assumed to only be visible in the first case due to its anisotropic nature and small opening angle \citep[$\sim5\degr$, e.g.][]{Metzger12}.

\subsection{Accretion flow emission}
\label{sec:accretionflow_emission}

\subsubsection{Super-Eddington ZEBRA emission}

During the super-Eddington phase, the emitted spectrum of the ZEBRA is a single blackbody with bolometric luminosity equal to the Eddington luminosity of the SMBH, consistent with optical observations of TDEs \citep[e.g.][]{Gezari21}. From the photospheric radius and temperature derived from the numerical simulations, the luminosity per wavelength (given in cm here) is therefore
\begin{equation}
L_\lambda = 4\pi R_{\rm phot}^2 \frac{2\pi hc^2}{\lambda^5\left(e^{\frac{hc}{\lambda k_B T_{\rm phot}}} - 1\right)}{\rm erg\ s}^{-1} {\rm\ cm}^{-1}.
\label{eq:zebras_bb}
\end{equation}

\subsubsection{Transition between the super-Eddington and sub-Eddington phases}
\label{sec:collapse}

Accurately modelling the collapse of the ZEBRA to a thin disk is beyond the scope of this work. We therefore instead used a smoothed two component model, the final ZEBRA component derived in the super-Eddington phase and a thin disk component with emission modelled as in Section \ref{sec:thindisk}. We take $t_{\rm end}$ as the time at which $\dot{M}_{\rm acc}$ drops below the Eddington limit and $t_{\rm decay}$ as a decay timescale, taken to be two weeks here. The ZEBRA component, and in the on-axis case the accompanying jet component, is faded exponentially as $e^{-(t-t_{\rm end})/t_{\rm decay}}$ while the thin disk similarly rises as $1-e^{-(t-t_{\rm end})/t_{\rm decay}}$ and will rapidly start to dominate over the ZEBRA. The relatively quick timescale is motivated both by the sharp decline observed in the optical light curves of the jetted TDEs, particularly J2058+05, and directly calculating the Kelvin-Helmholtz timescale, $\tau_{\rm KH}$. This timescale is given by the ratio of the thermal energy of the flow, $E_{\rm th}\sim GM_{\rm BH}\mathscr{M}/\mathscr{R}$, to its luminosity, $L\sim4\pi GM_{\rm BH}c/\kappa$, i.e.
\begin{equation}
    \tau_{\rm KH} \sim \frac{E_{\rm th}}{L} \sim \frac{\kappa\mathscr{M}}{4\pi Rc}
\end{equation}
where assuming $\mathscr\sim 0.1$\msol, $R\sim10^{15}$ cm and $\kappa \sim 0.34$ cm$^2$ g$^{-1}$ as above yields a $\tau_{\rm KH}$ of approximately two days. This rapid cooling leads in turn to the rapid collapse of the ZEBRA. We note, however, that a change to the decay timescale would have little impact on the observable properties of such a TDE.

\subsubsection{Sub-Eddington thin disk phase}
\label{sec:thindisk}

The temperature of a geometrically thin and optically thick disk varies according to its radius and the accretion rate as:
\begin{equation}
T(R) = \left[\frac{3 G M_{\rm BH} \dot{M}_{\rm acc}}{8 \pi R^3\sigma_{\rm SB}}\left[1-\left(\frac{R_{\rm in}}{R}\right)^{1/2}\right]\right]^{1/4} {\rm K}
\end{equation}
where $\sigma_{\rm SB}$ is the Stefan-Boltzmann constant and $R_{\rm in}$ is the inner radius of the accretion disk, taken to be 3 times the Schwarzschild radius of the black hole (i.e. $R_{\rm in} = 3 R_S = 6 G M_{\rm BH} / c^2$). The short viscous timescale of the disk means $\dot{M}_{\rm acc} \sim \dot{M}_{\rm fb}$. We took the outer radius of the disk, $R_{\rm out}$ to be twice the radius of closest approach, i.e. for our chosen $\beta = 1$, $R_{\rm out} = 2 r_p = 2 r_t$. To account for the varying temperature across the disk, we use the standard procedure of treating the disk as a series of concentric annuli \citep{Frank02}. Each of these annuli emits as a blackbody with temperature determined by its central radius and a luminosity per unit wavelength (cm) of
\begin{equation}
L_\lambda = 2\pi R_{\rm annulus}dR \frac{2\pi hc^2}{\lambda^5\left(e^{\frac{hc}{\lambda k_B T_{\rm annulus}}} - 1\right)} {\rm erg\ s}^{-1} {\rm\ cm}^{-1}
\label{eq:zebras_disk}
\end{equation}
where $R_{\rm annulus}$ is the central radius of the annulus, $R_{\rm annulus} = R_{\rm inner} + \frac{dR}{2}$, and $dR$ is the width of the annulus, $R_{\rm outer} - R_{\rm inner}$. Integrating over the annuli will therefore give the total emission from the disk.

Both how the ZEBRA collapses and the orientation of the disrupted star's orbit with respect to the SMBH's spin are likely to significantly affect the resultant thin disk. Along with the Lense-Thirring effect, these result in a high probability that the disk will be `tilted' or `twisted' \citep[e.g.][]{Ivanov18,Raj21}. The resulting precession of the disk leads to variability in the observed emission \citep[e.g.][]{Stone12,Franchini16}. While we do not model this variability here, we note that it is likely to have a non-negligible but small impact on the final light curve.

\subsection{Jet emission}
\label{sec:jet_emission}

The jet emission is modelled following \citet{Metzger12}, \citet{Berger12}, \citet{Zauderer13} and \citet{Eftekhari18} who all examine \textit{Swift} J1644+57's jet. We therefore assumed, at least intially, that any jet contribution at optical wavelengths would be dominated by synchrotron emission. We also assumed that the total luminosity is equally split between two jets and therefore the kinetic luminosity of each is $L_{j,\text{iso}}/2$.

Synchrotron spectra are characterised by power law segments between break frequencies, $\nu_{\text{a}}$, the self-absorption frequency; $\nu_{\text{m}}$, the peak frequency; and $\nu_{\text{c}}$, the cooling frequency. The photon index, $\Gamma$, varies between these segments and at higher frequencies is also dependent on the electron energy distribution. This is assumed to be a power law distribution with index $p$ and to properly derive the spectrum, both $p$ and the break frequencies need to be calculated.

Examination of J1644+57's X-ray light curve indicated the kinetic luminosity of the jet remained constant for $\sim10^6$ s. This luminosity was therefore used to determine the synchrotron parameters in the modelling of \citet{Metzger12}, \citet{Berger12}, \citet{Zauderer13} and \citet{Eftekhari18} as any additional energy injected at later times is relatively negligible compared to that injected here. For our model, we used the peak kinetic luminosity from Figure 1 of \citetalias{Wu18} the time of which was used to define $t_{pk}$. Note that \citetalias{Wu18} also include a radiative efficiency of $\varepsilon = 0.1$.

A suitable value for $p$ can be determined from observations of J1644+57, J2058+05 and J1112-82. We initially assume that all emission, from X-ray to radio, was dominated by synchrotron. The \textit{Swift} BAT and XRT spectra of both J1644+57 and J2058+05 are reasonably well fit at early times (days) with a power law with $\Gamma\sim1.6$ \citep{Levan11,Cenko12} while J1112-82 had a harder spectrum with $\Gamma\sim1.3$ \citep{Brown15}. There is some evidence of slight steepening in J2058+05's spectrum at $\sim200$ days to $\Gamma\sim1.8$ but it also becomes shallower again at $\Gamma\sim1.7$ at $\sim 1$ year. J1644+57 also shows evidence of later steepening \citep{Levan16} but it is after the drop in the X-ray light curve and is likely to be evidence of the disk component starting to dominate over the jet. From the X-ray, therefore we assumed $\Gamma = 1.6$ to be suitably representative of the jet's general behaviour and that it was in the regime $\nu_{\text{m}} < \nu < \nu_{\text{c}}$. The radio emission of J1644+57 has also been extensively investigated and modelled. While not directly fitted, it was found to be consistent with a $p$ in the range  2.3 to 2.5 \citep{Metzger12,Berger12,Zauderer13}.

Therefore, from the closure relations of \citet{Granot02} with $\Gamma = 1.6$ and the radio modelling of J1644+57, we assumed $p=2\Gamma-1 = 2.2$ to be accurate. For $p = 2.2$, the break frequencies are given by \citep{Granot02,Metzger12,Berger12,Zauderer13}:
\begin{equation}
   \nu_{\text{a}}(t) = \begin{cases} 5.5\times10^{9}\times\epsilon_{e,-1}^{-1}\epsilon_{B,-2}^{1/5}\\ \hspace{3em}\times L_{j,\text{iso},48}^{-2/5}t_{j,6}^{-1}n^{6/5}\left(\frac{t}{t_j}\right)^{-1}\text{Hz}, & t\leq t_{\text{pk}} \\[10pt]
    1.1\times10^{10}\times\epsilon_{e,-1}^{-1}\epsilon_{B,-2}^{1/5}\\ \hspace{3em}\times L_{j,\text{iso},48}^{-2/5}t_{j,6}^{-1}n^{6/5}\left(\frac{t}{t_j}\right)^{-3/5} \text{Hz}, & t> t_{\text{pk}} \end{cases}
\label{eq:zebras_jetnua}
\end{equation}
\begin{equation}
    \nu_{\text{m}}(t) =\begin{cases} 2.7\times10^{10} \times \epsilon_{e,-1}^{2}\epsilon_{B,-2}^{1/2}\\ \hspace{3em}\times  L_{j,\text{iso},48}^{1/2} t_{j,6}^{-1}\left(\frac{t}{t_j}\right)^{-1}\text{Hz}, & t\leq t_{\text{pk}} \\[10pt]
    7.6\times10^{10} \times \epsilon_{e,-1}^{2}\epsilon_{B,-2}^{1/2}\\ \hspace{3em}\times L_{j,\text{iso},48}^{1/2}t_{j,6}^{-1}\left(\frac{t}{t_j}\right)^{-3/2}\text{Hz}, & t>t_{\text{pk}} \end{cases}
\label{eq:zebras_jetnum}
\end{equation}
\begin{equation}
    \nu_{\text{c}}(t) = 2.9\times10^{14}\times\epsilon_{B,-2}^{-3/2}L_{j,\text{iso},48}^{1/2}t_{j,6}n^{-2}\left(\frac{t}{t_j}\right)^{1/2}\text{Hz for all }t
\label{eq:zebras_jetnuc}
\end{equation}
where we use the notation $X\equiv10^y X_y$; $\epsilon_e = 0.1$ and $\epsilon_B = 0.01$ are the fractions of the total internal energy directed into the electrons and magnetic field respectively; and $n_{18}$ is the circumnuclear density at a fiducial radius of $r=10^{18}$ cm. While there is evidence of $n_{18}$ to vary over time in the case of J1644+57 \citep{Berger12}, we assumed that this was either specific to J1644+57 or any variation would be negligible. We therefore assume $n_{18} = 1$ cm$^{-3}$ at all times. 

The luminosity normalisation of the spectrum can be derived similarly and we find
\begin{equation}
L_{\nu < \nu_{\text{a}}}(t) = \begin{cases} 20 \times \epsilon_{e,-1}L_{j,\text{iso},48}^{3/2}t_{j,6}^{2}n^{-3/2} \\ \hspace{3em}\times \theta_{j,-1}^{2}\nu_{10}^{2}\left(\frac{t}{t_j}\right)^{2}\text{erg s$^{-1}$ Hz$^{-1}$},  & t\leq t_{\text{pk}} \\[10pt] 175 \times \epsilon_{e,-1}L_{j,\text{iso},48}^{3/2}t_{j,6}^{2}n^{-3/2} \\ \hspace{3em}\times\theta_{j,-1}^{2}\nu_{10}^{2}\left(\frac{t}{t_j}\right)^{1/2}\text{erg s$^{-1}$ Hz$^{-1}$}, & t>t_{\text{pk}}\end{cases}
\label{eq:zebras_jetL}
\end{equation}
where $\theta_{j} = 0.1$ is the opening angle of the jet.

\section{Results}
\label{sec:Results}

\subsection{Off-axis light curve}

\begin{figure}
 \includegraphics[width=\columnwidth]{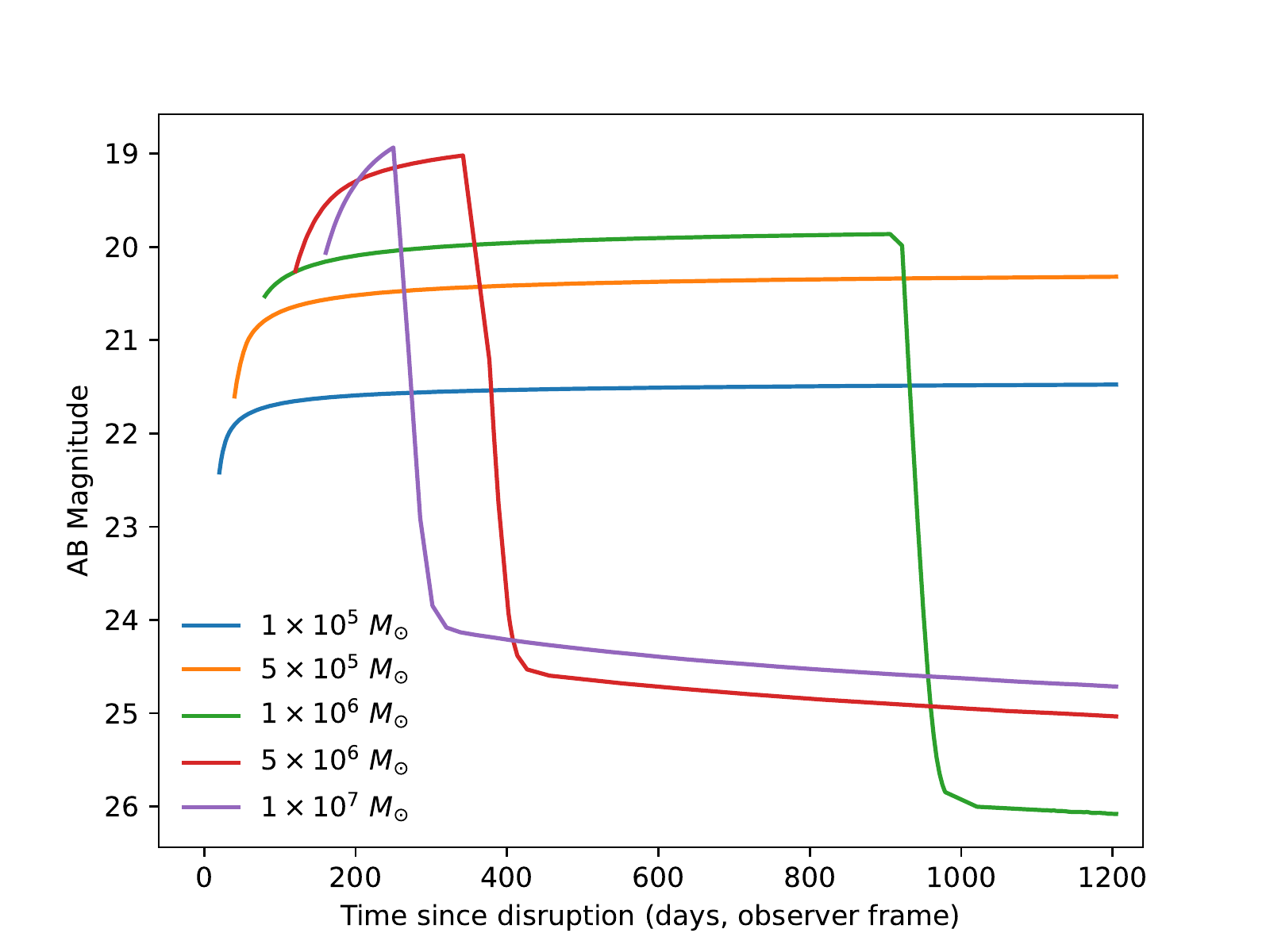}
 \caption{The observed off-axis $g$ band light curves of TDEs with varying SMBH mass from $10^5-10^7$ \msol. All sources are at a redshift of $z=0.1$.}
 \label{fig:gband}
\end{figure}

Due to the anisotropic nature of the jet, the sole component visible in the off-axis light curve is the ZEBRA flow and the later thin disk. The resultant SDSS \textit{g} band light curves for each SMBH mass are shown in Figure \ref{fig:gband}. All sources are taken to have a redshift of $z=0.1$, comparable to the peak of the distribution of observed TDEs' redshifts \citep{Qin22}.

The distinct phases of the TDEs' behaviour are easy to identify in the light curves. The super-Eddington phase begins with a fairly rapid rise before slowing and starting to plateau. The length of the super-Eddington phase has an obvious SMBH mass dependence with higher SMBH masses resulting in a shorter timescale.  For the $1\times10^5$ and $5\times10^5$ \msol\ cases, this timescale was actually longer than the $\sim3$ years the simulations covered. The TDE is relatively bright during this phase especially for higher SMBH masses, peaking at $\sim19$ mags for the $1\times10^7$ \msol. Following the super-Eddington phase, the collapse phase and sub-Eddington phases are also distinct. In the sub-Eddington phase, the decay was consistent with $t^{-5/12}$ as expected for a thin disk model at optical wavelengths \citep[e.g.][]{Lodato11,Gezari21}. However, this phase is significantly fainter than the super-Eddington phase by $\sim 5 - 6$ magnitudes.

\begin{figure}
 \includegraphics[width=\columnwidth]{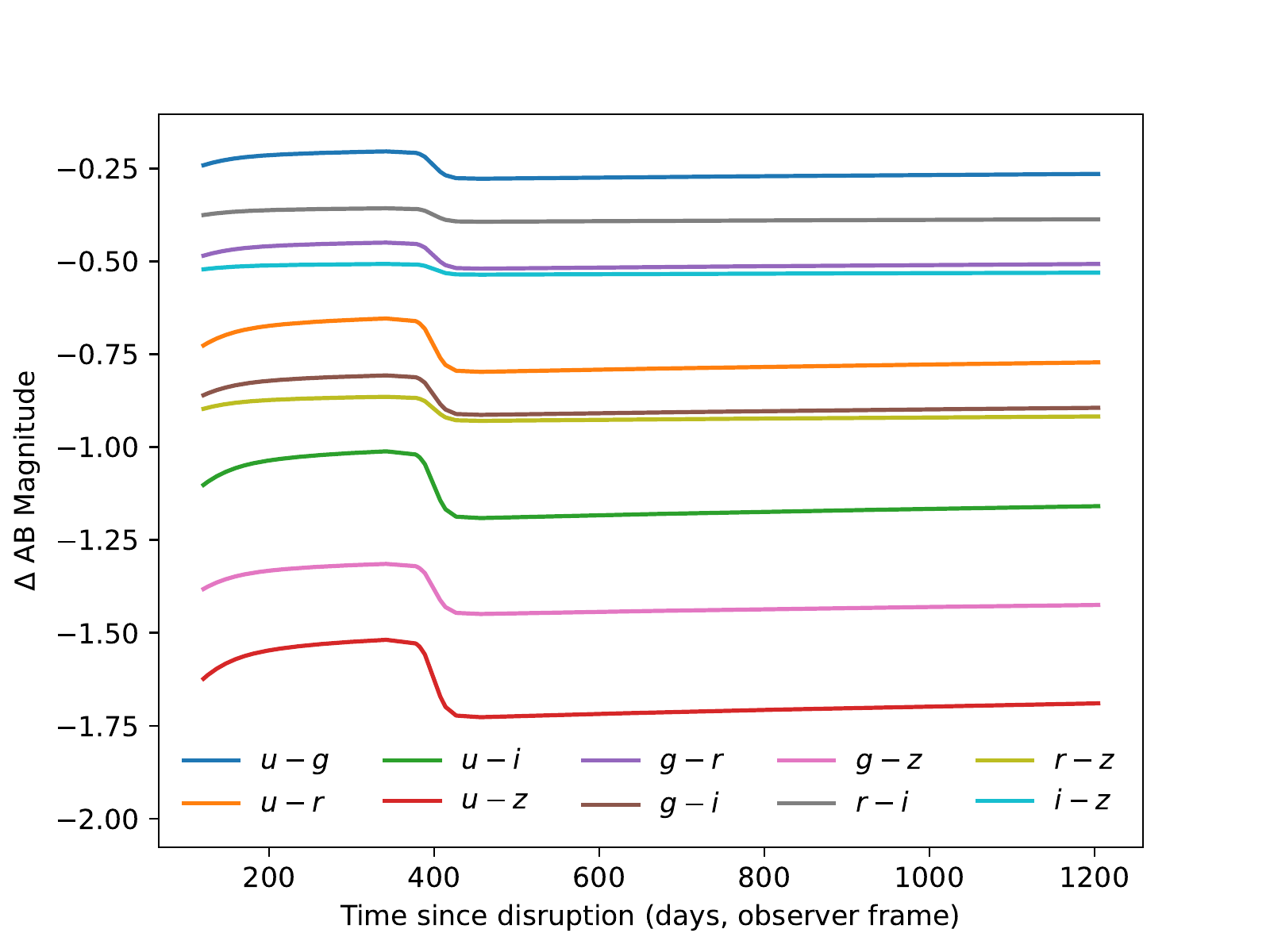}
 \caption{The colour of a TDE at a $5\times10^6$ \msol\ SMBH and at a redshift of $z=0.1$ for each combination of the SDSS $ugriz$ bands. All magnitudes are in the observed band.}
 \label{fig:colour}
\end{figure}

We also investigated the colour change of the transient as shown in Figure \ref{fig:colour} using the SDSS $ugriz$ bands. Generally, the TDE's colours are relatively constant in time particularly within the individual phases. There is a more significant shift during the transition between the super-Eddington and sub-Eddington phases, but this is still small with a maximum change of $\lesssim0.25$ mags and would therefore require precise photometry to measure. Observationally, therefore, the end of the super-Eddington phase would be best identified using the drop in the light curve, as shown in Figure \ref{fig:gband}, rather than the spectral behaviour of the transient.

\subsection{On-axis light curve}

\begin{figure}
 \includegraphics[width=\columnwidth]{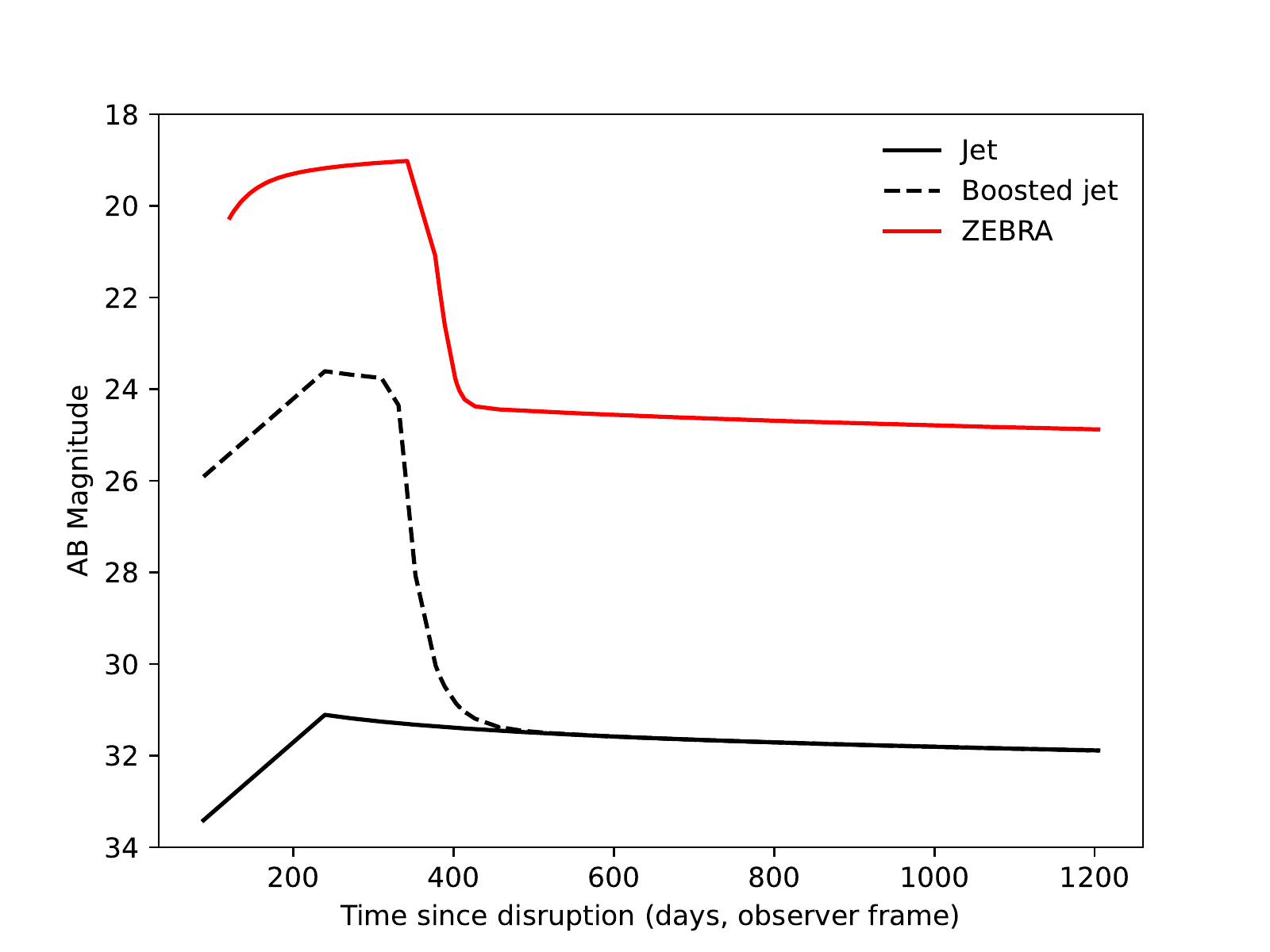}
 \caption{A comparison of the expected contribution from the ZEBRA (red) and jet (black solid) components to the on-axis $g$ band light curve for a TDE at a $5\times10^6$ \msol\ SMBH and at a redshift of $z=0.1$. The boosted jet due to synchrotron self-Compton emission is also shown in the dashed line.}
 \label{fig:jet}
\end{figure}

In addition to the accretion flow, the on-axis case positions the observer such that the jet contribution is fully visible. Adding the jet contribution to the off-axis light curve therefore produces the on-axis light curve. However, as shown in Figure \ref{fig:jet}, the jet's synchrotron contribution at optical wavelengths is $\sim4.8$ orders of magnitude fainter than that of the ZEBRA at the jet's peak time and $\sim2.8$ orders of magnitude fainter during the sub-Eddington thin disk phase. The jet's synchrotron emission therefore has a negligible effect on the overall optical light curve.

However, there is evidence that the dominant emission mechanism at optical wavelengths for the observed jetted TDEs was not synchrotron \citep[e.g.][]{Bloom11,Lu16,Crumley16}. We therefore examined the possibilities of other mechanisms such as synchrotron self-Compton or external inverse-Compton emission. Such mechanisms could boost the jet's optical luminosity by up to $\sim3$ orders of magnitude during the super-Eddington phase. We include such a boosted jet in Figure \ref{fig:jet} by increasing the jet's optical luminosity by a factor of 1000 and adding an exponential decline from when the acccretion rate drops below the Eddington limit. However, we found that even with this increase, the jet contribution is still $\sim1.8$ orders of magnitude fainter than the ZEBRA at the jet's peak and remains $\sim2.8$ orders of magnitude fainter during the sub-Eddington phase. The jet consequently remains unlikely to make any significant impact on the light curve.

In addition to any jet contribution, it is possible that the emission observed from the ZEBRA flow itself may be different when observed on-axis. The density of the ZEBRA varies according to angle and therefore the optical depth can also vary. For an on-axis observer, therefore, it might be possible to view inside the `funnel' of the flow and the inner regions of the ZEBRA to dominate the observed emission rather than the surface. In this case, however, we have assumed the jet effectively fills the funnel and that its outer sheath is optically thick \citep{Coughlin20a}. Any emisison from the inner regions is therefore blocked and only the ZEBRA's surface emission is observed.

In this version of the ZEBRA model, therefore, the on-axis and off-axis are functionally identical during the super-Eddington phase with only a small increase in luminosity ($\sim 15\%$) during the sub-Eddington thin disk phase for the on-axis model.

\section{Comparison to observed TDEs}
\label{sec:Discussion}

\begin{figure}
 \includegraphics[width=\columnwidth]{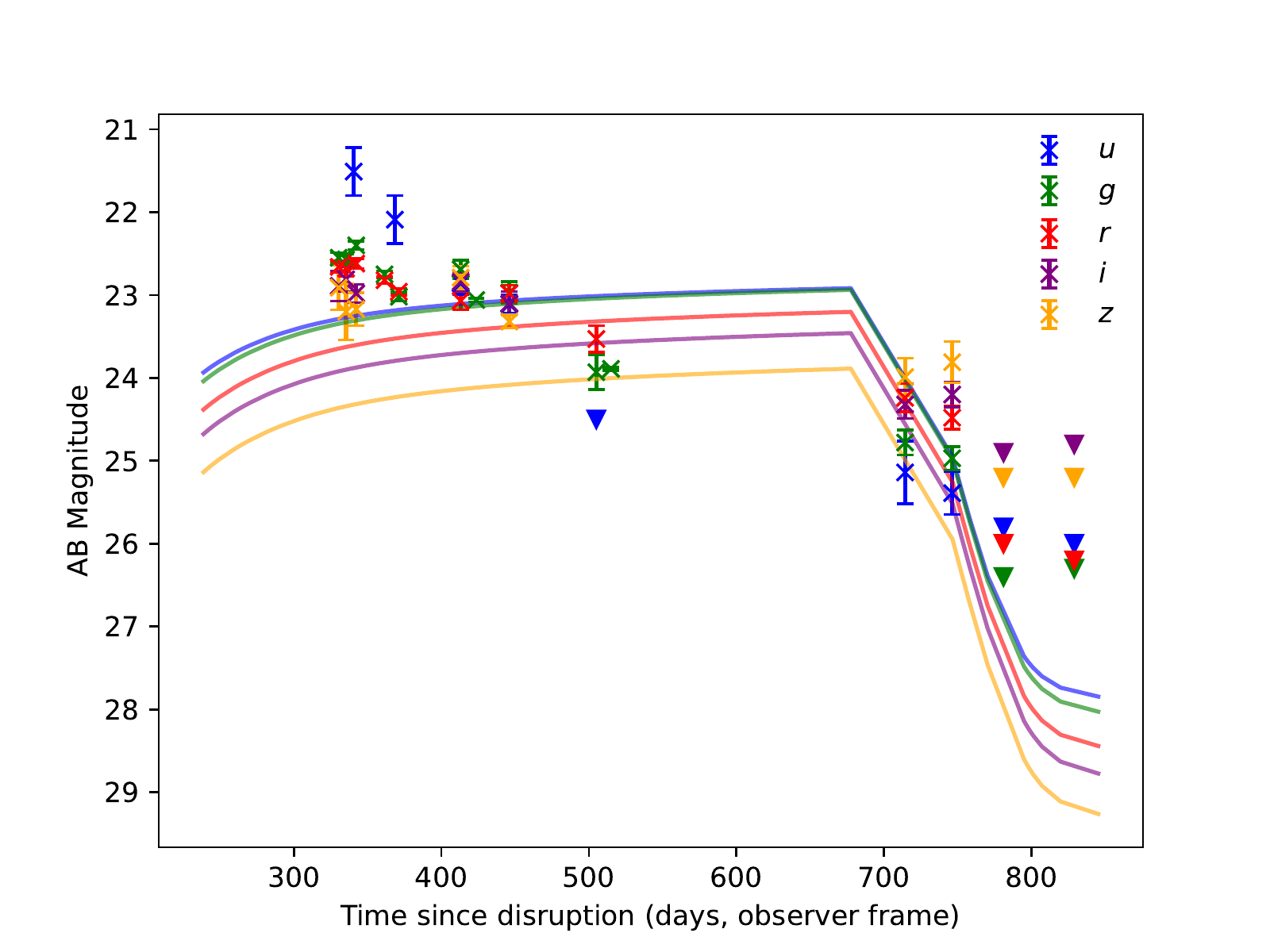}
 \caption{The optical light curve of Swift J2058+05 and the ZEBRA model light curves of a TDE at a $5\times10^6$ \msol\ SMBH at a redshift of $z=1.1853$ with a Milky Way extinction of $E(B-V) = 0.095$ mag and a host extinction of $A_V = 0.2$ mag. Note that this is in the observer frame, hence the model light curves reach the end of the super-Eddington phase later than in Figure \ref{fig:gband}. All magnitudes are in the observed band.}
 \label{fig:2058}
\end{figure}

To examine how accurately the ZEBRA model captures the behaviour of real events, we now compare the model light curves to the data collected for J2058+05. Of the jetted TDEs, J2058+05 has the best sampled optical light curve, shown in Figure \ref{fig:2058}. Its inferred SMBH mass is $\sim5\times10^6$ \msol \citep{Cenko12,Pasham15} and we also include the ZEBRA model light curves for such a TDE at the same redshift, applying extinctions of $E(B-V) = 0.095$ mag for the Milky Way \citep{Schlafly11} and $A_V = 0.2$ mag for J2058+05's host \citep{Pasham15}.

There are some commonalities between the model and the observed light curves. In particular, the timescale on which J2058+05's optical light curve starts to drop much more steeply is consistent with the collapse phase being reached in the model. The X-ray light curve \citep[e.g. Figure 4 in][]{Zauderer13} also drops at a similar time reinforcing the view that the time at which super-Eddington accretion ends is accurately captured by the ZEBRA model. The luminosity of the model and observed light curves are also broadly comparable, particularly in the $g$ band.

However, there are also significant differences. In particular, J2058+05 is somewhat brighter than the model at early times, particularly in the higher energy bands, and fades prior to the sudden drop rather than continuing to rise as in the model. The colour changes much more significantly than in the model and J2058+05 cools much more quickly than expected for the ZEBRA model.

\begin{figure}
 \includegraphics[width=\columnwidth]{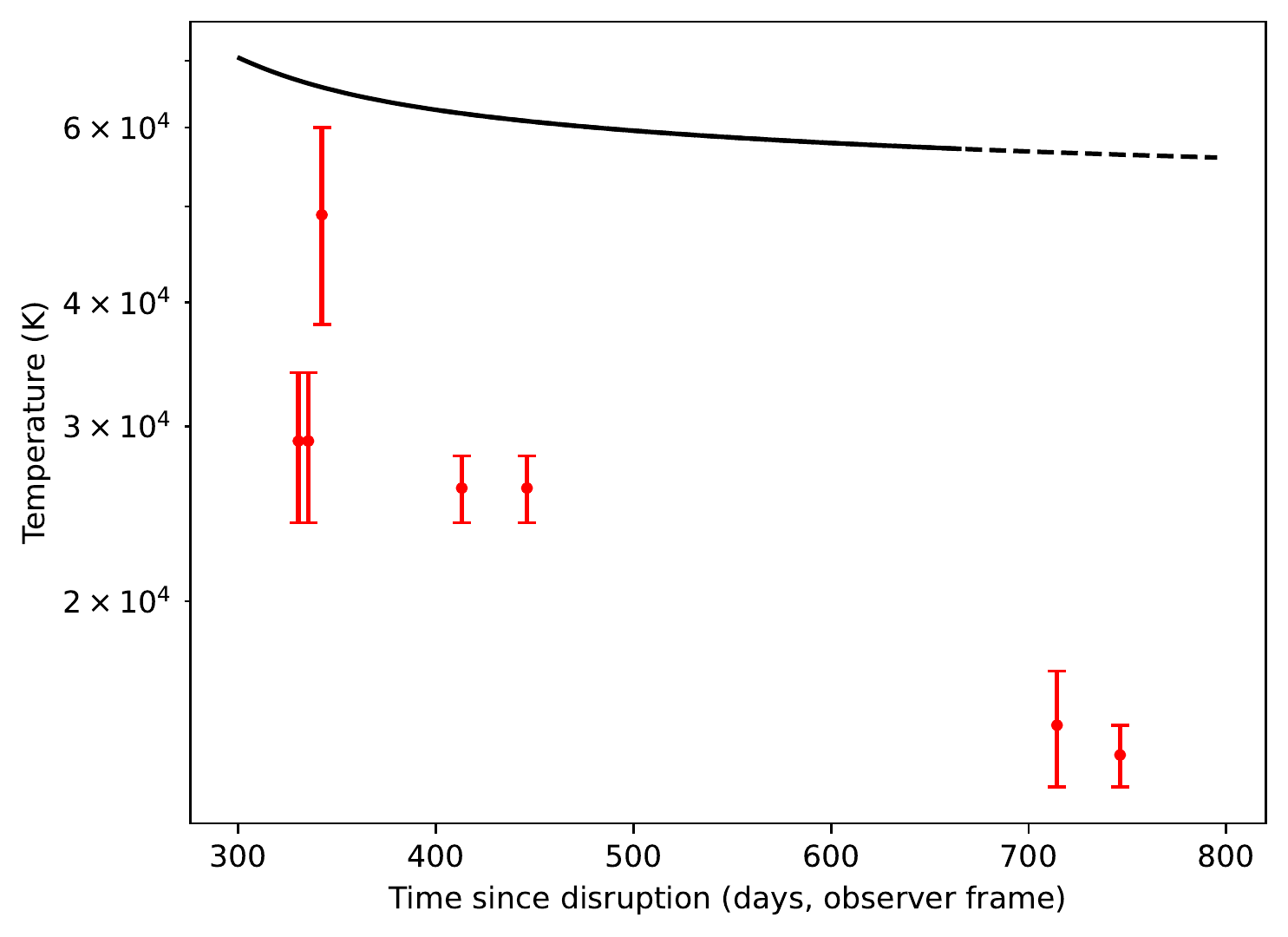}
 \includegraphics[width=\columnwidth]{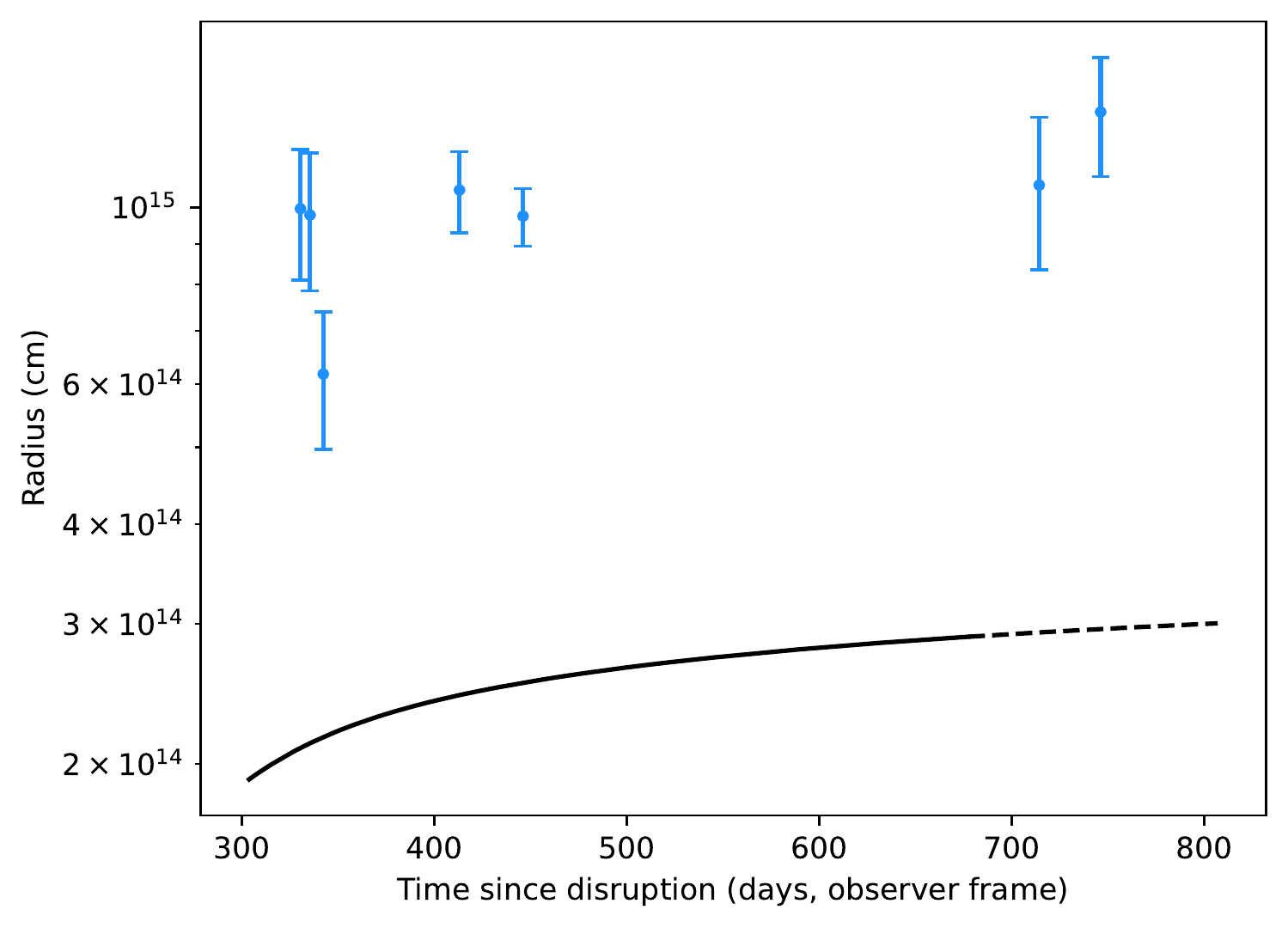}
 \includegraphics[width=\columnwidth]{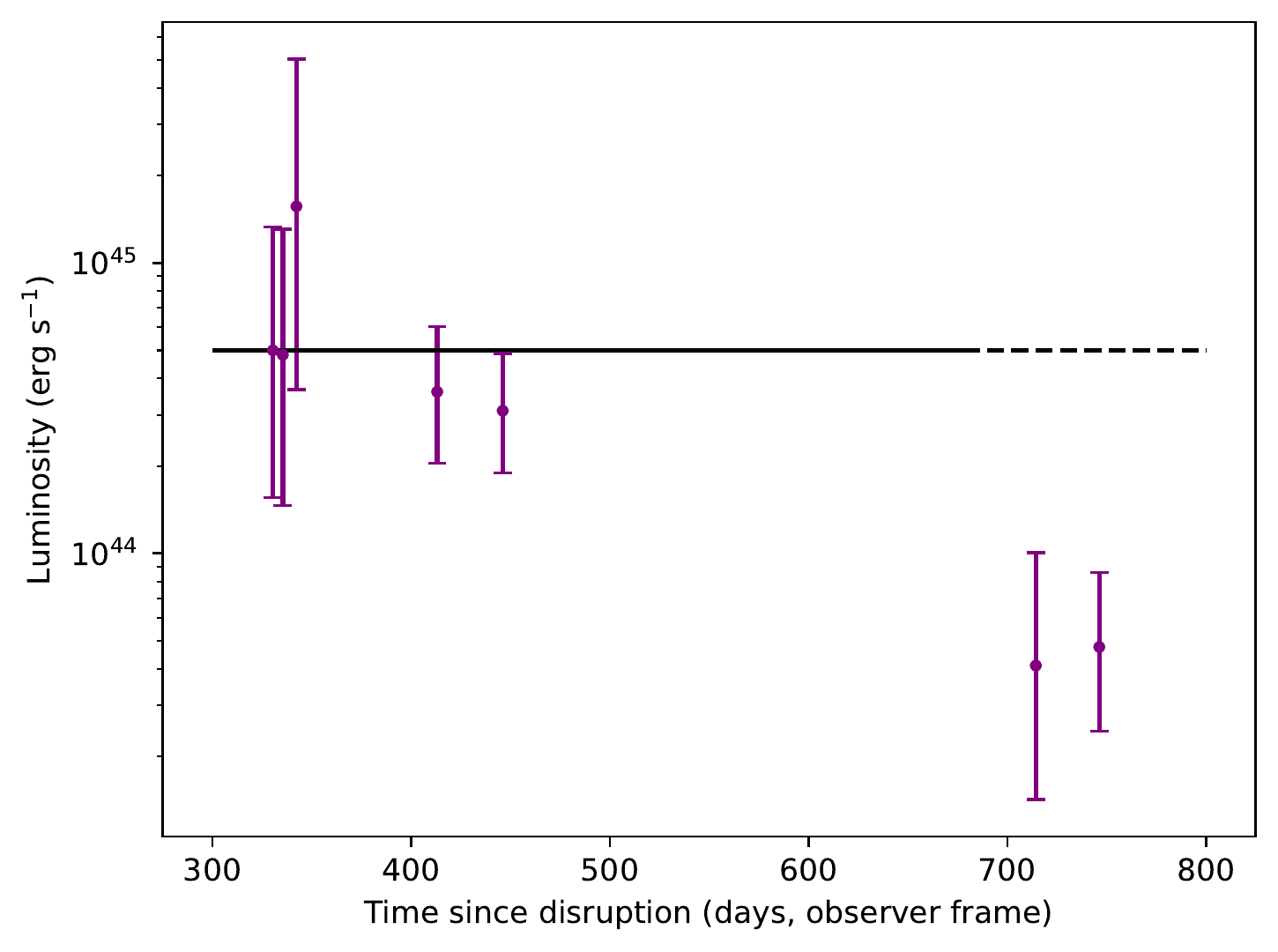}
 \caption{The blackbody temperature (top), radius (middle) and bolometric luminosity (bottom) measured by \citet{Pasham15} for J2058+05. The black lines indicate the expected behaviour from the ZEBRA model \citepalias{Wu18}. The transition to dashed lines indicate the accretion rate becoming sub-Eddington.}
 \label{fig:bbcomp}
\end{figure}

\citet{Pasham15} examined the UVOIR SED of J2058+05 finding it be reasonably well fit with a single blackbody. In Figure \ref{fig:bbcomp}, we compare the parameters inferred from their fits with those derived using the ZEBRA model by \citetalias{Wu18}. We found that J2058+05's bolometric luminosity is generally consistent with the Eddington luminosity of a $5\times10^6$ \msol\ SMBH as predicted by the ZEBRA model. However, the radius of the fitted blackbody is $\sim2.7 - 4.1$ times larger while the fitted temperature is $\sim1.4 - 2.3$ times smaller than that expected for a ZEBRA. The accretion flow is therefore likely to be significantly more inflated than suggested by the model. It should be noted, however, that this discrepancy between observed blackbody radius and that predicted by theory is shared by many models of TDEs \citep[e.g.][]{Gezari21}. J1644+57 and J1112-82 behave similarly to J2058+05, although have poorer sampled optical light curves, and the discrepancies identified here also apply to them.

While other TDEs are not as clearly super-Eddington as the jetted events, it is still valuable to compare to a larger sample such as that of \citet{vanVelzen21}. Their sample had peak blackbody luminosities of $\sim10^{43.6}$ to $\sim10^{44.6}$ erg s$^{-1}$, consistent with the Eddington luminosities of SMBHs in the range $\sim10^{5.6}$ to $\sim10^{6.6}$ \msol. However, they predict only a weak correlation between peak luminosity and BH mass, indicating the luminosity may not be limited by the Eddington limit. In terms of temperature, there are significant differences around peak times with observed temperatures up to a factor of $\sim$three smaller than predicted by the ZEBRA model. In most cases, the temperature was also found to be rising rather than falling as in the ZEBRA's super-Eddington phase. \citet{Nicholl22} also analyse the luminosity and temperature behaviour of a sample of 32 TDEs, including those of \citet{vanVelzen21}. Their findings are similar, with bolometric luminosities varying as a few tenths of the Eddington luminosity, and although their temperatures are still rising, they are generally more in agreement with those of the ZEBRA model. The shape of the light curves in both samples also differ significantly, with much faster rises and slower decays observed by both \citeauthor{vanVelzen21} and \citeauthor{Nicholl22} than predicted. In particular, there is no evidence for the extreme decline as the ZEBRA collapses. Overall, similarly to the jetted TDEs, there does appear to be some agreement in terms of luminosity but many other properties are discrepant. Generally these discrepancies are more extreme than is seen for jetted TDEs, however, this is not unexpected as many of the TDEs in both samples are likely to be sub-Eddington and therefore subject to a different behavioural regime.

It is also worth noting that the ZEBRA models presented here (taken from \citetalias{Wu18}) do not cover the full range of possible dynamics, and therefore the full range of possible fallback rates, in a TDE. For example, the TDEs explored by \citetalias{Wu18} have the pericentre distance of the stellar orbit equal to the tidal radius and it is known that partial disruptions can produce fallback rates that initial rise more steeply and then decay more steeply at late times \citep[see e.g.][]{Nixon21}. Similarly the simulations in \citetalias{Wu18} employ a $\gamma=5/3$ polytrope to model the star; this is an excellent approximation to the properties of  some low-mass stars, but different fallback rates may be produced by different types of stars \citep[see e.g.][]{Golightly19b} and partial TDEs \citep{Guillochon13, Coughlin19}. While it is beyond the scope of the current paper to investigate such effects, this would be an interesting avenue for further exploration.

\section{Conclusions}
\label{sec:Conclusions}

We have examined the ZEBRA model of tidal disruption events and derived optical light curves for a star with solar mass and radius disrupted by SMBHs of various masses. These light curves indicate a long-lived plateau phase during the period of super-Eddington accretion with a timescale inversely related to the mass of the SMBH. This is followed by a collapse to a more typical thin disk as the accretion rate drops below the Eddington limit. We have also investigated the potential contribution to the optical light curve by a jet but found it to be negligible compared to the emission from the accretion flow itself.

Comparing these results to the jetted TDE J2058+05, we found that some properties were consistent, such as the timescale for which the event remained super-Eddington and the bolometric luminosity. However, other properties, such as the rise times and blackbody radius and temperature, were found to differ between the model and the observed data. The rise time for the ZEBRA (i.e. the formation of the flow) is not self-consistently accounted for in the model (see the discussion in \citetalias{Coughlin&Begelman}). It is possible that a more realistic description of the flow would be to have the average Bernoulli parameter be zero, but to allow there to be a finite (but small) gradient in the Bernoulli parameter throughout the flow, with the inner regions being bound (and liberating some of that energy as it accretes), and the outer envelope having a region of positive Bernoulli parameter. This would then lead to a small outward velocity, which would expand the envelope to larger radii and likely bring the model predictions into better agreement with observations. In addition, incorporating the radiation-dominated outflow solutions of \citet{Coughlin20a} could aid in understanding the nature of jetted emission and any contribution made at optical wavelengths.

Overall, we find that while the ZEBRA model does not fully explain the behaviour of observed TDEs, it is consistent with many aspects. Further refinements to the model could assist in resolving the remaining discrepancies and, with the wealth of forthcoming data from large optical surveys and transient follow-up, help to answer the question of exactly what powers super-Eddington TDEs.

\section*{Acknowledgements}

RAJEF acknowledges support from the Science and Technology Facilities Council, the UK Space Agency and the European Union’s Horizon 2020 Programme under the AHEAD2020 project (grant agreement number 871158). CJN acknowledges support from the Science and Technology Facilities Council [grant number ST/W000857/1], and the Leverhulme Trust (grant number RPG-2021-380). ERC~acknowledges support from the National Science Foundation through grant AST-2006684, and a Ralph E.~Powe Junior Faculty Enhancement Award through the Oakridge Associated Universities. We also thank Gavin Lamb, Mike Goad and Mat Page for helpful discussion and suggestions, and the anonymous referee for their useful feedback.


\section*{Data Availability}

The data from \citetalias{Wu18} used to generate the light curves in this paper were obtained by private communication. The model light curves will be shared on reasonable request to the corresponding author.


\bibliographystyle{mnras}
\bibliography{TDEs,software,071227}

\begin{thebibliography}{}
\makeatletter
\relax
\def\mn@urlcharsother{\let\do\@makeother \do\$\do\&\do\#\do\^\do\_\do\%\do\~}
\def\mn@doi{\begingroup\mn@urlcharsother \@ifnextchar [ {\mn@doi@}
  {\mn@doi@[]}}
\def\mn@doi@[#1]#2{\def\@tempa{#1}\ifx\@tempa\@empty \href
  {http://dx.doi.org/#2} {doi:#2}\else \href {http://dx.doi.org/#2} {#1}\fi
  \endgroup}
\def\mn@eprint#1#2{\mn@eprint@#1:#2::\@nil}
\def\mn@eprint@arXiv#1{\href {http://arxiv.org/abs/#1} {{\tt arXiv:#1}}}
\def\mn@eprint@dblp#1{\href {http://dblp.uni-trier.de/rec/bibtex/#1.xml}
  {dblp:#1}}
\def\mn@eprint@#1:#2:#3:#4\@nil{\def\@tempa {#1}\def\@tempb {#2}\def\@tempc
  {#3}\ifx \@tempc \@empty \let \@tempc \@tempb \let \@tempb \@tempa \fi \ifx
  \@tempb \@empty \def\@tempb {arXiv}\fi \@ifundefined
  {mn@eprint@\@tempb}{\@tempb:\@tempc}{\expandafter \expandafter \csname
  mn@eprint@\@tempb\endcsname \expandafter{\@tempc}}}

\bibitem[\protect\citeauthoryear{{Andalman}, {Liska}, {Tchekhovskoy},
  {Coughlin}  \& {Stone}}{{Andalman} et~al.}{2022}]{Andalman22}
{Andalman} Z.~L.,  {Liska} M. T.~P.,  {Tchekhovskoy} A.,  {Coughlin} E.~R.,
  {Stone} N.,  2022, \mn@doi [\mnras] {10.1093/mnras/stab3444}, \href
  {https://ui.adsabs.harvard.edu/abs/2022MNRAS.510.1627A} {510, 1627}

\bibitem[\protect\citeauthoryear{{Begelman}}{{Begelman}}{1978}]{Begelman78}
{Begelman} M.~C.,  1978, \mn@doi [\mnras] {10.1093/mnras/184.1.53}, \href
  {https://ui.adsabs.harvard.edu/abs/1978MNRAS.184...53B} {184, 53}

\bibitem[\protect\citeauthoryear{{Berger}, {Zauderer}, {Pooley}, {Soderberg},
  {Sari}, {Brunthaler}  \& {Bietenholz}}{{Berger} et~al.}{2012}]{Berger12}
{Berger} E.,  {Zauderer} A.,  {Pooley} G.~G.,  {Soderberg} A.~M.,  {Sari} R.,
  {Brunthaler} A.,   {Bietenholz} M.~F.,  2012, \mn@doi [\apj]
  {10.1088/0004-637X/748/1/36}, \href
  {https://ui.adsabs.harvard.edu/abs/2012ApJ...748...36B} {748, 36}

\bibitem[\protect\citeauthoryear{{Blandford} \& {Znajek}}{{Blandford} \&
  {Znajek}}{1977}]{Blandford77}
{Blandford} R.~D.,  {Znajek} R.~L.,  1977, \mn@doi [\mnras]
  {10.1093/mnras/179.3.433}, \href
  {https://ui.adsabs.harvard.edu/abs/1977MNRAS.179..433B} {179, 433}

\bibitem[\protect\citeauthoryear{{Bloom} et~al.,}{{Bloom}
  et~al.}{2011}]{Bloom11}
{Bloom} J.~S.,  et~al., 2011, \mn@doi [Science] {10.1126/science.1207150},
  \href {https://ui.adsabs.harvard.edu/abs/2011Sci...333..203B} {333, 203}

\bibitem[\protect\citeauthoryear{{Bonnerot} \& {Lu}}{{Bonnerot} \&
  {Lu}}{2020}]{Bonnerot20}
{Bonnerot} C.,  {Lu} W.,  2020, \mn@doi [\mnras] {10.1093/mnras/staa1246},
  \href {https://ui.adsabs.harvard.edu/abs/2020MNRAS.495.1374B} {495, 1374}

\bibitem[\protect\citeauthoryear{{Bonnerot}, {Price}, {Lodato}  \&
  {Rossi}}{{Bonnerot} et~al.}{2017}]{Bonnerot17}
{Bonnerot} C.,  {Price} D.~J.,  {Lodato} G.,   {Rossi} E.~M.,  2017, \mn@doi
  [\mnras] {10.1093/mnras/stx1210}, \href
  {https://ui.adsabs.harvard.edu/abs/2017MNRAS.469.4879B} {469, 4879}

\bibitem[\protect\citeauthoryear{{Brown}, {Levan}, {Stanway}, {Tanvir},
  {Cenko}, {Berger}, {Chornock}  \& {Cucchiaria}}{{Brown}
  et~al.}{2015}]{Brown15}
{Brown} G.~C.,  {Levan} A.~J.,  {Stanway} E.~R.,  {Tanvir} N.~R.,  {Cenko}
  S.~B.,  {Berger} E.,  {Chornock} R.,   {Cucchiaria} A.,  2015, \mn@doi
  [\mnras] {10.1093/mnras/stv1520}, \href
  {https://ui.adsabs.harvard.edu/abs/2015MNRAS.452.4297B} {452, 4297}

\bibitem[\protect\citeauthoryear{{Burrows} et~al.,}{{Burrows}
  et~al.}{2011}]{Burrows11}
{Burrows} D.~N.,  et~al., 2011, \mn@doi [\nat] {10.1038/nature10374}, \href
  {https://ui.adsabs.harvard.edu/#abs/2011Natur.476..421B} {476, 421}

\bibitem[\protect\citeauthoryear{{Cannizzo}, {Lee}  \& {Goodman}}{{Cannizzo}
  et~al.}{1990}]{Cannizzo90}
{Cannizzo} J.~K.,  {Lee} H.~M.,   {Goodman} J.,  1990, \mn@doi [\apj]
  {10.1086/168442}, \href
  {https://ui.adsabs.harvard.edu/abs/1990ApJ...351...38C} {351, 38}

\bibitem[\protect\citeauthoryear{{Cenko} et~al.,}{{Cenko}
  et~al.}{2012}]{Cenko12}
{Cenko} S.~B.,  et~al., 2012, \mn@doi [\apj] {10.1088/0004-637X/753/1/77},
  \href {https://ui.adsabs.harvard.edu/#abs/2012ApJ...753...77C} {753}

\bibitem[\protect\citeauthoryear{{Coughlin} \& {Begelman}}{{Coughlin} \&
  {Begelman}}{2014}]{Coughlin&Begelman}
{Coughlin} E.~R.,  {Begelman} M.~C.,  2014, \mn@doi [\apj]
  {10.1088/0004-637X/781/2/82}, \href
  {https://ui.adsabs.harvard.edu/#abs/2014ApJ...781...82C} {781}

\bibitem[\protect\citeauthoryear{{Coughlin} \& {Begelman}}{{Coughlin} \&
  {Begelman}}{2020}]{Coughlin20a}
{Coughlin} E.~R.,  {Begelman} M.~C.,  2020, \mn@doi [\mnras]
  {10.1093/mnras/staa3026}, \href
  {https://ui.adsabs.harvard.edu/abs/2020MNRAS.499.3158C} {499, 3158}

\bibitem[\protect\citeauthoryear{{Coughlin} \& {Nixon}}{{Coughlin} \&
  {Nixon}}{2019}]{Coughlin19}
{Coughlin} E.~R.,  {Nixon} C.~J.,  2019, \mn@doi [\apjl]
  {10.3847/2041-8213/ab412d}, \href
  {https://ui.adsabs.harvard.edu/abs/2019ApJ...883L..17C} {883, L17}

\bibitem[\protect\citeauthoryear{{Coughlin}, {Nixon}, {Begelman}  \&
  {Armitage}}{{Coughlin} et~al.}{2016}]{Coughlin16b}
{Coughlin} E.~R.,  {Nixon} C.,  {Begelman} M.~C.,   {Armitage} P.~J.,  2016,
  \mn@doi [\mnras] {10.1093/mnras/stw770}, \href
  {http://adsabs.harvard.edu/abs/2016MNRAS.459.3089C} {459, 3089}

\bibitem[\protect\citeauthoryear{{Crumley}, {Lu}, {Santana}, {Hern{\'a}ndez},
  {Kumar}  \& {Markoff}}{{Crumley} et~al.}{2016}]{Crumley16}
{Crumley} P.,  {Lu} W.,  {Santana} R.,  {Hern{\'a}ndez} R.~A.,  {Kumar} P.,
  {Markoff} S.,  2016, \mn@doi [\mnras] {10.1093/mnras/stw967}, \href
  {https://ui.adsabs.harvard.edu/abs/2016MNRAS.460..396C} {460, 396}

\bibitem[\protect\citeauthoryear{{Dai}, {McKinney}, {Roth}, {Ramirez-Ruiz}  \&
  {Miller}}{{Dai} et~al.}{2018}]{Dai18}
{Dai} L.,  {McKinney} J.~C.,  {Roth} N.,  {Ramirez-Ruiz} E.,   {Miller} M.~C.,
  2018, \mn@doi [\apjl] {10.3847/2041-8213/aab429}, \href
  {https://ui.adsabs.harvard.edu/abs/2018ApJ...859L..20D} {859, L20}

\bibitem[\protect\citeauthoryear{{Eftekhari}, {Berger}, {Zauderer}, {Margutti}
  \& {Alexander}}{{Eftekhari} et~al.}{2018}]{Eftekhari18}
{Eftekhari} T.,  {Berger} E.,  {Zauderer} B.~A.,  {Margutti} R.,   {Alexander}
  K.~D.,  2018, \mn@doi [\apj] {10.3847/1538-4357/aaa8e0}, \href
  {https://ui.adsabs.harvard.edu/abs/2018ApJ...854...86E} {854, 86}

\bibitem[\protect\citeauthoryear{{Franchini}, {Lodato}  \&
  {Facchini}}{{Franchini} et~al.}{2016}]{Franchini16}
{Franchini} A.,  {Lodato} G.,   {Facchini} S.,  2016, \mn@doi [\mnras]
  {10.1093/mnras/stv2417}, \href
  {https://ui.adsabs.harvard.edu/abs/2016MNRAS.455.1946F} {455, 1946}

\bibitem[\protect\citeauthoryear{{Frank}, {King}  \& {Raine}}{{Frank}
  et~al.}{2002}]{Frank02}
{Frank} J.,  {King} A.,   {Raine} D.~J.,  2002, {Accretion Power in
  Astrophysics: Third Edition}.
Cambridge University Press

\bibitem[\protect\citeauthoryear{{Gafton} \& {Rosswog}}{{Gafton} \&
  {Rosswog}}{2011}]{Gafton11}
{Gafton} E.,  {Rosswog} S.,  2011, \mn@doi [\mnras]
  {10.1111/j.1365-2966.2011.19528.x}, \href
  {https://ui.adsabs.harvard.edu/abs/2011MNRAS.418..770G} {418, 770}

\bibitem[\protect\citeauthoryear{{Gezari}}{{Gezari}}{2021}]{Gezari21}
{Gezari} S.,  2021, \mn@doi [\araa] {10.1146/annurev-astro-111720-030029},
  \href {https://ui.adsabs.harvard.edu/abs/2021ARA&A..59...21G} {59, 21}

\bibitem[\protect\citeauthoryear{{Golightly}, {Nixon}  \&
  {Coughlin}}{{Golightly} et~al.}{2019}]{Golightly19b}
{Golightly} E.~C.~A.,  {Nixon} C.~J.,   {Coughlin} E.~R.,  2019, \mn@doi
  [\apjl] {10.3847/2041-8213/ab380d}, \href
  {https://ui.adsabs.harvard.edu/abs/2019ApJ...882L..26G} {882, L26}

\bibitem[\protect\citeauthoryear{{Granot} \& {Sari}}{{Granot} \&
  {Sari}}{2002}]{Granot02}
{Granot} J.,  {Sari} R.,  2002, \mn@doi [\apj] {10.1086/338966}, \href
  {https://ui.adsabs.harvard.edu/#abs/2002ApJ...568..820G} {568, 820}

\bibitem[\protect\citeauthoryear{{Guillochon} \& {McCourt}}{{Guillochon} \&
  {McCourt}}{2017}]{Guillochon17b}
{Guillochon} J.,  {McCourt} M.,  2017, \mn@doi [\apj]
  {10.3847/2041-8213/834/2/L19}, \href
  {https://ui.adsabs.harvard.edu/abs/2017ApJ...834L..19G} {834, L19}

\bibitem[\protect\citeauthoryear{{Guillochon} \& {Ramirez-Ruiz}}{{Guillochon}
  \& {Ramirez-Ruiz}}{2013}]{Guillochon13}
{Guillochon} J.,  {Ramirez-Ruiz} E.,  2013, \mn@doi [\apj]
  {10.1088/0004-637X/767/1/25}, \href
  {https://ui.adsabs.harvard.edu/#abs/2013ApJ...767...25G} {767}

\bibitem[\protect\citeauthoryear{{Ivanov}, {Zhuravlev}  \&
  {Papaloizou}}{{Ivanov} et~al.}{2018}]{Ivanov18}
{Ivanov} P.~B.,  {Zhuravlev} V.~V.,   {Papaloizou} J.~C.~B.,  2018, \mn@doi
  [\mnras] {10.1093/mnras/sty2493}, \href
  {https://ui.adsabs.harvard.edu/abs/2018MNRAS.481.3470I} {481, 3470}

\bibitem[\protect\citeauthoryear{{Kasen} \& {Ramirez-Ruiz}}{{Kasen} \&
  {Ramirez-Ruiz}}{2010}]{Kasen10}
{Kasen} D.,  {Ramirez-Ruiz} E.,  2010, \mn@doi [\apj]
  {10.1088/0004-637X/714/1/155}, \href
  {https://ui.adsabs.harvard.edu/abs/2010ApJ...714..155K} {714, 155}

\bibitem[\protect\citeauthoryear{{Levan} et~al.,}{{Levan}
  et~al.}{2011}]{Levan11}
{Levan} A.~J.,  et~al., 2011, \mn@doi [Science] {10.1126/science.1207143},
  \href {https://ui.adsabs.harvard.edu/#abs/2011Sci...333..199L} {333, 199}

\bibitem[\protect\citeauthoryear{{Levan}, {Crowther}, {de Grijs}, {Langer},
  {Xu}  \& {Yoon}}{{Levan} et~al.}{2016}]{Levan16}
{Levan} A.,  {Crowther} P.,  {de Grijs} R.,  {Langer} N.,  {Xu} D.,   {Yoon}
  S.-C.,  2016, \mn@doi [\ssr] {10.1007/s11214-016-0312-x}, \href
  {http://adsabs.harvard.edu/abs/2016SSRv..202...33L} {202, 33}

\bibitem[\protect\citeauthoryear{{Lodato} \& {Rossi}}{{Lodato} \&
  {Rossi}}{2011}]{Lodato11}
{Lodato} G.,  {Rossi} E.~M.,  2011, \mn@doi [\mnras]
  {10.1111/j.1365-2966.2010.17448.x}, \href
  {http://adsabs.harvard.edu/abs/2011MNRAS.410..359L} {410, 359}

\bibitem[\protect\citeauthoryear{{Loeb} \& {Ulmer}}{{Loeb} \&
  {Ulmer}}{1997}]{Loeb97}
{Loeb} A.,  {Ulmer} A.,  1997, \mn@doi [\apj] {10.1086/304814}, \href
  {https://ui.adsabs.harvard.edu/#abs/1997ApJ...489..573L} {489, 573}

\bibitem[\protect\citeauthoryear{{Lu} \& {Kumar}}{{Lu} \& {Kumar}}{2016}]{Lu16}
{Lu} W.,  {Kumar} P.,  2016, \mn@doi [\mnras] {10.1093/mnras/stv2324}, \href
  {https://ui.adsabs.harvard.edu/abs/2016MNRAS.458.1071L} {458, 1071}

\bibitem[\protect\citeauthoryear{{Metzger}}{{Metzger}}{2022}]{Metzger22}
{Metzger} B.~D.,  2022, \mn@doi [\apjl] {10.3847/2041-8213/ac90ba}, \href
  {https://ui.adsabs.harvard.edu/abs/2022ApJ...937L..12M} {937, L12}

\bibitem[\protect\citeauthoryear{{Metzger} \& {Stone}}{{Metzger} \&
  {Stone}}{2016}]{Metzger16}
{Metzger} B.~D.,  {Stone} N.~C.,  2016, \mn@doi [\mnras]
  {10.1093/mnras/stw1394}, \href
  {https://ui.adsabs.harvard.edu/abs/2016MNRAS.461..948M} {461, 948}

\bibitem[\protect\citeauthoryear{{Metzger}, {Giannios}  \& {Mimica}}{{Metzger}
  et~al.}{2012}]{Metzger12}
{Metzger} B.~D.,  {Giannios} D.,   {Mimica} P.,  2012, \mn@doi [\mnras]
  {10.1111/j.1365-2966.2011.20273.x}, \href
  {https://ui.adsabs.harvard.edu/abs/2012MNRAS.420.3528M} {420, 3528}

\bibitem[\protect\citeauthoryear{{Miles}, {Coughlin}  \& {Nixon}}{{Miles}
  et~al.}{2020}]{Miles20}
{Miles} P.~R.,  {Coughlin} E.~R.,   {Nixon} C.~J.,  2020, \mn@doi [\apj]
  {10.3847/1538-4357/ab9c9f}, \href
  {https://ui.adsabs.harvard.edu/abs/2020ApJ...899...36M} {899, 36}

\bibitem[\protect\citeauthoryear{{Nicholl}, {Lanning}, {Ramsden}, {Mockler},
  {Lawrence}, {Short}  \& {Ridley}}{{Nicholl} et~al.}{2022}]{Nicholl22}
{Nicholl} M.,  {Lanning} D.,  {Ramsden} P.,  {Mockler} B.,  {Lawrence} A.,
  {Short} P.,   {Ridley} E.~J.,  2022, \mn@doi [\mnras]
  {10.1093/mnras/stac2206}, \href
  {https://ui.adsabs.harvard.edu/abs/2022MNRAS.515.5604N} {515, 5604}

\bibitem[\protect\citeauthoryear{{Nixon}, {Coughlin}  \& {Miles}}{{Nixon}
  et~al.}{2021}]{Nixon21}
{Nixon} C.~J.,  {Coughlin} E.~R.,   {Miles} P.~R.,  2021, \mn@doi [\apj]
  {10.3847/1538-4357/ac1bb8}, \href
  {https://ui.adsabs.harvard.edu/abs/2021ApJ...922..168N} {922, 168}

\bibitem[\protect\citeauthoryear{{Norman}, {Nixon}  \& {Coughlin}}{{Norman}
  et~al.}{2021}]{Norman21}
{Norman} S.~M.~J.,  {Nixon} C.~J.,   {Coughlin} E.~R.,  2021, \mn@doi [\apj]
  {10.3847/1538-4357/ac2ee8}, \href
  {https://ui.adsabs.harvard.edu/abs/2021ApJ...923..184N} {923, 184}

\bibitem[\protect\citeauthoryear{{Pasham} et~al.,}{{Pasham}
  et~al.}{2015}]{Pasham15}
{Pasham} D.~R.,  et~al., 2015, \mn@doi [\apj] {10.1088/0004-637X/805/1/68},
  \href {https://ui.adsabs.harvard.edu/abs/2015ApJ...805...68P} {805, 68}

\bibitem[\protect\citeauthoryear{{Phinney}}{{Phinney}}{1989}]{Phinney89}
{Phinney} E.~S.,  1989, in {Morris} M.,  ed., The Center of the Galaxy.
  Proceedings of the 136th Symposium of the International Astronomical Union.
p.~543

\bibitem[\protect\citeauthoryear{{Price} et~al.,}{{Price}
  et~al.}{2018}]{Price17}
{Price} D.~J.,  et~al., 2018, \mn@doi [\pasa] {10.1017/pasa.2018.25}, \href
  {https://ui.adsabs.harvard.edu/abs/2018PASA...35...31P} {35, e031}

\bibitem[\protect\citeauthoryear{{Qin} et~al.,}{{Qin} et~al.}{2022}]{Qin22}
{Qin} Y.-J.,  et~al., 2022, \mn@doi [\apjs] {10.3847/1538-4365/ac2fa1}, \href
  {https://ui.adsabs.harvard.edu/abs/2022ApJS..259...13Q} {259, 13}

\bibitem[\protect\citeauthoryear{{Raj}, {Nixon}  \& {Do{\u{g}}an}}{{Raj}
  et~al.}{2021}]{Raj21}
{Raj} A.,  {Nixon} C.~J.,   {Do{\u{g}}an} S.,  2021, \mn@doi [\apj]
  {10.3847/1538-4357/abdc24}, \href
  {https://ui.adsabs.harvard.edu/abs/2021ApJ...909...81R} {909, 81}

\bibitem[\protect\citeauthoryear{{Rees}}{{Rees}}{1988}]{Rees88}
{Rees} M.~J.,  1988, \mn@doi [\nat] {10.1038/333523a0}, \href
  {https://ui.adsabs.harvard.edu/#abs/1988Natur.333..523R} {333, 523}

\bibitem[\protect\citeauthoryear{{Schlafly} \& {Finkbeiner}}{{Schlafly} \&
  {Finkbeiner}}{2011}]{Schlafly11}
{Schlafly} E.~F.,  {Finkbeiner} D.~P.,  2011, \mn@doi [\apj]
  {10.1088/0004-637X/737/2/103}, \href
  {https://ui.adsabs.harvard.edu/#abs/2011ApJ...737..103S} {737, 103}

\bibitem[\protect\citeauthoryear{{S{\k{a}}dowski}, {Tejeda}, {Gafton},
  {Rosswog}  \& {Abarca}}{{S{\k{a}}dowski} et~al.}{2016}]{Sadowski16}
{S{\k{a}}dowski} A.,  {Tejeda} E.,  {Gafton} E.,  {Rosswog} S.,   {Abarca} D.,
  2016, \mn@doi [\mnras] {10.1093/mnras/stw589}, \href
  {https://ui.adsabs.harvard.edu/abs/2016MNRAS.458.4250S} {458, 4250}

\bibitem[\protect\citeauthoryear{{Steinberg} \& {Stone}}{{Steinberg} \&
  {Stone}}{2022}]{Steinberg22}
{Steinberg} E.,  {Stone} N.~C.,  2022, arXiv e-prints, \href
  {https://ui.adsabs.harvard.edu/abs/2022arXiv220610641S} {p. arXiv:2206.10641}

\bibitem[\protect\citeauthoryear{{Stone} \& {Loeb}}{{Stone} \&
  {Loeb}}{2012}]{Stone12}
{Stone} N.,  {Loeb} A.,  2012, \mn@doi [\prl] {10.1103/PhysRevLett.108.061302},
  \href {https://ui.adsabs.harvard.edu/abs/2012PhRvL.108f1302S} {108, 061302}

\bibitem[\protect\citeauthoryear{{Strubbe} \& {Quataert}}{{Strubbe} \&
  {Quataert}}{2009}]{Strubbe09}
{Strubbe} L.~E.,  {Quataert} E.,  2009, \mn@doi [\mnras]
  {10.1111/j.1365-2966.2009.15599.x}, \href
  {https://ui.adsabs.harvard.edu/#abs/2009MNRAS.400.2070S} {400, 2070}

\bibitem[\protect\citeauthoryear{{Tchekhovskoy}, {Metzger}, {Giannios}  \&
  {Kelley}}{{Tchekhovskoy} et~al.}{2014}]{Tchekohovskoy14}
{Tchekhovskoy} A.,  {Metzger} B.~D.,  {Giannios} D.,   {Kelley} L.~Z.,  2014,
  \mn@doi [\mnras] {10.1093/mnras/stt2085}, \href
  {https://ui.adsabs.harvard.edu/abs/2014MNRAS.437.2744T} {437, 2744}

\bibitem[\protect\citeauthoryear{{Ulmer}}{{Ulmer}}{1999}]{Ulmer99}
{Ulmer} A.,  1999, \mn@doi [\apj] {10.1086/306909}, \href
  {https://ui.adsabs.harvard.edu/abs/1999ApJ...514..180U} {514, 180}

\bibitem[\protect\citeauthoryear{{Wiersema} et~al.,}{{Wiersema}
  et~al.}{2012}]{Wiersema12}
{Wiersema} K.,  et~al., 2012, \mn@doi [\mnras]
  {10.1111/j.1365-2966.2011.20379.x}, \href
  {https://ui.adsabs.harvard.edu/abs/2012MNRAS.421.1942W} {421, 1942}

\bibitem[\protect\citeauthoryear{{Wiersema} et~al.,}{{Wiersema}
  et~al.}{2020}]{Wiersema20}
{Wiersema} K.,  et~al., 2020, \mn@doi [\mnras] {10.1093/mnras/stz3106}, \href
  {https://ui.adsabs.harvard.edu/abs/2020MNRAS.491.1771W} {491, 1771}

\bibitem[\protect\citeauthoryear{{Wu}, {Coughlin}  \& {Nixon}}{{Wu}
  et~al.}{2018}]{Wu18}
{Wu} S.,  {Coughlin} E.~R.,   {Nixon} C.,  2018, \mn@doi [\mnras]
  {10.1093/mnras/sty971}, \href
  {https://ui.adsabs.harvard.edu/abs/2018MNRAS.478.3016W} {478, 3016}

\bibitem[\protect\citeauthoryear{{Zauderer} et~al.,}{{Zauderer}
  et~al.}{2011}]{Zauderer11}
{Zauderer} B.~A.,  et~al., 2011, \mn@doi [\nat] {10.1038/nature10366}, \href
  {https://ui.adsabs.harvard.edu/#abs/2011Natur.476..425Z} {476, 425}

\bibitem[\protect\citeauthoryear{{Zauderer}, {Berger}, {Margutti}, {Pooley},
  {Sari}, {Soderberg}, {Brunthaler}  \& {Bietenholz}}{{Zauderer}
  et~al.}{2013}]{Zauderer13}
{Zauderer} B.~A.,  {Berger} E.,  {Margutti} R.,  {Pooley} G.~G.,  {Sari} R.,
  {Soderberg} A.~M.,  {Brunthaler} A.,   {Bietenholz} M.~F.,  2013, \mn@doi
  [\apj] {10.1088/0004-637X/767/2/152}, \href
  {https://ui.adsabs.harvard.edu/abs/2013ApJ...767..152Z} {767, 152}

\bibitem[\protect\citeauthoryear{{van Velzen} et~al.,}{{van Velzen}
  et~al.}{2011}]{vanVelzen11}
{van Velzen} S.,  et~al., 2011, \mn@doi [\apj] {10.1088/0004-637X/741/2/73},
  \href {https://ui.adsabs.harvard.edu/#abs/2011ApJ...741...73V} {741}

\bibitem[\protect\citeauthoryear{{van Velzen}, {Frail}, {K{\"o}rding}  \&
  {Falcke}}{{van Velzen} et~al.}{2013}]{vanVelzen13}
{van Velzen} S.,  {Frail} D.~A.,  {K{\"o}rding} E.,   {Falcke} H.,  2013,
  \mn@doi [\aap] {10.1051/0004-6361/201220426}, \href
  {https://ui.adsabs.harvard.edu/abs/2013A&A...552A...5V} {552, A5}

\bibitem[\protect\citeauthoryear{{van Velzen} et~al.,}{{van Velzen}
  et~al.}{2021}]{vanVelzen21}
{van Velzen} S.,  et~al., 2021, \mn@doi [\apj] {10.3847/1538-4357/abc258},
  \href {https://ui.adsabs.harvard.edu/abs/2021ApJ...908....4V} {908, 4}

\makeatother
\end{thebibliography}



\appendix


\bsp	
\label{lastpage}
\end{document}